\documentclass[useAMS,usenatbib]{./mn2e}
\usepackage{times}
\usepackage{aas_macros}
\usepackage{amssymb}
\usepackage{rotating}
\usepackage{lscape}
\usepackage{multirow}
\usepackage{longtable}
\usepackage{natbib}
\def\type{png}

\def\lcdm{$\Lambda$CDM }
\def\orca{{\tt ORCA} }
\def\mass{{\it h}^{-1}M_{\odot}}
\def\mpc{{\it h}^{-1}\rm Mpc}
\def\nmin{$\rm N_{\rm min}$}
\def\cm{c_{\rm m20}}
\def\probthresh{\rm P_{\rm thresh}}
\def\sigmacrit{$\Sigma_{\rm crit}$}
\def\grtsim{\mathrel{\hbox{\rlap{\hbox{\lower2pt\hbox{$\sim$}}}\raise2pt\hbox{$>$}}}}
\def\lesssim{\mathrel{\hbox{\rlap{\hbox{\lower2pt\hbox{$\sim$}}}\raise2pt\hbox{$<$}}}}

\begin{document}

\title[ORCA]{ORCA: The Overdense Red-sequence Cluster Algorithm}

\author[Murphy et al.] {\parbox[h]{\textwidth}{D.~N.~A.~Murphy$^{1}$\thanks{E-mail:
    david.murphy@durham.ac.uk}, J.~E.~Geach$^{1,2}$, R.\ G.\ Bower$^{1}$}
  \vspace*{3pt}\\
\noindent $^{1}$ Institute for Computational Cosmology, Durham University, South Road, 
Durham. DH1 3LE. UK.\\
\noindent $^{2}$ Department of Physics, McGill University, Ernest Rutherford Building, 3600 Rue University, Montr\'eal, Qu\'ebec, Canada, H3A 2T8\\}

\date{}

\pagerange{\pageref{firstpage}--\pageref{lastpage}}\pubyear{2011}

\maketitle

\label{firstpage}

\begin{abstract} We present a new cluster detection algorithm designed for the
Panoramic Survey Telescope and Rapid Response System (Pan-STARRS)
survey but with generic application to any multiband data. The method
makes no prior assumptions about the properties of clusters other than
(a) the similarity in colour of cluster galaxies (the ``red
sequence'') and (b) an enhanced projected surface density. The
detector has three main steps: (i) it identifies cluster members by
photometrically filtering the input catalogue to isolate galaxies in
colour-magnitude space, (ii) a Voronoi diagram identifies regions of
high surface density, (iii) galaxies are grouped into clusters with a
Friends-of-Friends technique. Where multiple colours are available, we
require systems to exhibit sequences in two colours. In this paper we
present the algorithm and demonstrate it on two datasets. The first is
a 7 square degree sample of the deep Sloan Digital Sky Survey
equatorial stripe (Stripe 82), from which we detect 97 clusters with
$z \leq 0.6$. Benefiting from deeper data, we are 100\% complete in
the {\tt maxBCG} optically-selected cluster catalogue (based on
shallower single epoch SDSS data) and find an additional 78 previously
unidentified clusters. The second dataset is a mock Medium Deep Survey
(MDS) Pan-STARRS catalogue, based on the \lcdm model and a
semi-analytic galaxy formation recipe. Knowledge of galaxy-halo
memberships in the mock allows a quantification of algorithm
performance. We detect 305 mock clusters in haloes with mass
$>10^{13}\mass$ at $z\lesssim$0.6 and determine a spurious detection
rate of $<1\%$, consistent with tests on the Stripe 82 catalogue. The
detector performs well in the recovery of model \lcdm clusters. At the
median redshift of the catalogue, the algorithm achieves $>75\%$
completeness down to halo masses of $10^{13.4}\mass$ and recovers
$>75\%$ of the total stellar mass of clusters in haloes down to
$10^{13.8}\mass$. A companion paper \citep*[][hereafter
GMB11]{GMB2010} presents the complete cluster catalogue over the full
270\,deg$^2$ Stripe 82 catalogue.
\end{abstract}

\begin{keywords}
catalogues -- galaxies: clusters: general -- cosmology: observations
-- cosmology: large-scale structure of universe
\vspace*{-0.5 truecm}
 \end{keywords}

\section{Introduction}
 Galaxy clusters are integral tools in our drive to test the \lcdm
 cosmological model and our understanding of galaxy formation. The
 evolution of the cluster population with redshift for example, can
 impose important constraints on the matter density of the universe
 \citep{1996ApJ...462...32C,1997MNRAS.292..289E,2003A&A...402...53S}
 and the growth of primordial density fluctuations
 \citep{1990ApJ...351...10F,1993MNRAS.262.1023W,2009MNRAS.397.1125F}. The
 deep potential wells of clusters offer a suite of laboratories within
 which detailed studies of gas-galaxy interactions are possible. There
 is evidence that clusters have been in place for a significant
 fraction of the star-forming history of the universe, meaning they
 can provide a unique insight into the how environmental effects shape
 the evolutionary path of galaxies.

 The cluster mass budget is dominated by the presence of dark matter
 \citep[$\sim$85\%, for a comprehensive review
   see][]{2005RvMP...77..207V}, making them ideal sites for identifying
 strongly-lensed background galaxies \citep[][]{2007ApJ...654L..33S}
 and thus provide glimpses of the early star-forming universe
 \citep{2010Natur.464..733S}. Weak lensing studies can determine the
 projected mass distribution of clusters
 \citep[e.g.,][]{2004AJ....127.2544S} and in some cases the dark
 matter itself \citep{2006ApJ...648L.109C}. Hot intracluster gas also
 leaves an imprint on the cosmic microwave background (CMB) by
 way of the Sunyaev-Zel'dovich
 \citep[SZ,][]{1980ARA&A..18..537S,2002ARA&A..40..643C} effect via the
 inverse Compton scattering of CMB photons. At the megaparsec scale,
 clusters act as high-mass lamp-posts between the filamentary
 connected structure tracing out the cosmic web (\citeauthor{2004MNRAS.347..137P} \citeyear{2004MNRAS.347..137P};
 \citeauthor{2005MNRAS.359..272C} \citeyear{2005MNRAS.359..272C};
 \citeauthor*{MEF2010} \citeyear{MEF2010}).

There is therefore great merit in producing a homogeneous cluster
census of the Universe, and much effort has gone into producing
comprehensive cluster surveys. Efforts to this end are broadly
separated into two wavelength domains: the optical--near-IR and
X-ray. We note in passing that cluster detection by SZ-decrement in
the microwave is an emerging cluster survey technique that holds
promise at high redshift
\citep[][]{2009MNRAS.399L..84M,2010ApJ...721...90B,2010ApJS..191..423H,2010ApJ...722.1180V}.

X-ray detections exploit the hot intracluster gas accounting for the
bulk of the cluster baryonic mass component
\citep{1976A&A....49..137C,2002MNRAS.334L..11A}. X-ray selected
cluster catalogues tend to be robust to projection effects, probe
large volumes and produce a cluster sample with well characterised
masses. Cluster catalogues from large area X-ray surveys
\citep[e.g.,][]{1998MNRAS.301..881E} identify bright, massive
clusters, with their deep potential wells establishing the high
electron densities required for strong X-ray emission. Whilst a
cursory glance in the X-ray unveils the presence of high mass systems,
to select those with lower masses, unresolved gas components, distant
or gas-poor clusters, one must look to alternative approaches.

There has been a half-century history of cluster identification in the
optical regime. Early `eyeball' surveys of photographic plates
produced the earliest cluster catalogues
\citep{1958ApJS....3..211A,1961cgcg.book.....Z,1989ApJS...70....1A}
and allowed the first statistical study of the cluster
population. When cluster and group samples were later constructed with
the help of digitised photographic plates \citep[such as the
  APM;][]{1992ApJ...390L...1D} and galaxy spectra
\citep[][]{2004MNRAS.348..866E}, the task of identification passed
from human to machine. With the advent of wide-field multi-band CCD
imaging, assembly of vast galaxy samples has become the standard. For
example, Sloan Digital Sky Survey \citep[SDSS;][]{2000AJ....120.1579Y}
optical imaging data has vastly increased both the volume and detail
of detected astronomical sources, to date generating five-band {\it
  ugriz} photometry for $\sim 230$ million objects
\citep[DR7,][]{2009ApJS..182..543A}. Although one can estimate galaxy
redshifts photometrically based on SED template fitting
\citep{2003AJ....125..580C}, neural networks
\citep{2004PASP..116..345C} or a combination of the two
\citep[\S4.6]{2009ApJS..182..543A}, photo-{\it z}s are prone to large
uncertainties and are generally unsuitable for accurate 3D
reconstructions of the galaxy distribution \citep[although for recent
  approaches using the entire photometric redshift distribution,
  see][]{2008ApJ...681.1046L}.

Armed with only the angular positions of galaxies, automated
algorithms have been developed to identify clusters as projected
overdensities in the plane of the sky
\citep[][]{1996AJ....112.2454L,1996AJ....111..615P}. These often come
at the expense of model dependency and sensitivity to the boundaries
and holes common in galaxy catalogues. More geometric approaches have
made use of the Voronoi Tessellation (VT) to map the projected density
distribution of galaxies. Using the Voronoi cell area as a proxy for
the local galaxy density, VTs were first suggested as a non-parametric
means of astrophysical source detection by
\citet{1993PhRvE..47..704E}, and later cluster detection in
\citet{2001A&A...368..776R}. Voronoi techniques have also been used in
void detection \citep{1995ApJ...452...25R,1996ApJ...462L..13E} and the
identification of large scale structure
\citep{1991QJRAS..32...85I}. However, these approaches tend to suffer
from contamination arising from the inclusion of background and
foreground field galaxies.

\citet{2000AJ....120.2148G} proposed a powerful method that picks out
the near ubiquitous signature of galaxy clusters from photometric
surveys. Star formation rates of galaxies bound in the potential wells
of clusters are suppressed when the cold gas supply is depleted by
environmentally-driven stripping or starvation processes
\citep[][]{2000ApJ...540..113B}. The passively evolving stellar
populations in these galaxies develop strong metal absorption lines
blueward of 4000\AA\ giving rise to a break, or step, in their
spectra. In broad-band photometric filters, these cluster members
appear nearly uniformly red between the bands that straddle the
spectral break. Because cluster galaxies occupy a wide range of masses
(luminosities) these characteristic colours produce a distinct
ridgeline, or ``red sequence'' \citep[][]{1992MNRAS.254..601B} in
colour-magnitude space. The dichotomy between this quiescent
population of predominantly E/S0 galaxies and the star-forming
population of spiral-dominated field galaxies is observed as a
bi-modality of galaxy colours. With increasing redshift, the
4000\AA\ break moves redward; the \citet{2000AJ....120.2148G}
prescription for cluster detection exploits both the strong colour
bi-modality in the galaxy distribution, and the colour-redshift
relation to isolate clusters of galaxies over a range of epochs.

With a growing body of infrared data (specifically, the IRAC
cameras on-board the {\it Spitzer Space Telescope}), efforts such as
the {\it Spitzer} Adaptation of the Red-Sequence Cluster Survey
\citep[SpARCS,][]{2009ApJ...698.1943W} have already turned to pushing
red-sequence cluster searches beyond the optical/NIR regime. With
evidence of cluster sequences in place up to $z\sim1.5$
\citep{2010ApJ...716.1503P,2011MNRAS.tmp..778H} and perhaps as early
as $z=3$ \citep{2007MNRAS.377.1717K,2010A&A...509A..83D}, tracking the
4000\AA\ break further red-ward shows great potential in filling the
$1.4<z<2.2$ cluster desert. These distant systems may potentially hold
some crucial clues for our understanding of galaxy formation and
evolution.

Future observational campaigns such as the Large Synoptic Survey
Telescope \citep[LSST;][]{2008arXiv0805.2366I} are set to push forward
the frontiers of wide-area, deep multi-band optical sky surveys. More
immediately
Pan-STARRS-1\footnote{http://pan-starrs.ifa.hawaii.edu}\citep[PS-1;][]{2002SPIE.4836..154K},
the first of four 1.8m telescopes, is currently imaging 3/4 of the sky
with deep, and well characterised \citep{2010ApJS..191..376S}
five-band photometry. Algorithms capable of processing the
petabyte-scale sky surveys of these next-generation facilities will be
best placed to supply data products fully exploiting their
advances. Cluster selection by red sequence is set to remain highly
relevant to the construction of cluster catalogues using these
forthcoming surveys.

One approach to cluster detection in these deeper datasets is through
``matched-filter'' \citep[MF;][]{1996AJ....111..615P} algorithms that
distill the large body of collected cluster data into a likelihood
function, recovering systems by maximising the likelihood of survey
data fitting the model. In particular, these filters may specify the
cluster luminosity function, radial density distribution, behaviour of
the red sequence ridgeline and in some cases the presence of a central
Brightest Cluster Galaxy (BCG) \citep[{\tt
maxBCG};][]{2007ApJ...660..221K}. MF algorithms often confer redshift
and richness estimates as part of the detection procedure. The MF
technique has been successful in extracting cluster signals from a
diverse range of galaxy surveys, including the SDSS
\citep{2002AJ....123.1807G,2007ApJ...660..239K} and Canada France
Hawaii Telescope Legacy Survey
\citep[CFHTLS;][]{2005ApJS..157....1G,2009ApJ...706..571T}. The {\tt
maxBCG} SDSS cluster catalogue \citep{2007ApJ...660..239K} has
facilitated a more detailed study of the cluster red sequence
\citep{2009ApJ...702..745H}, which may in turn provide added
refinements to future algorithms.

However, the advantage of MF algorithms can also be their drawback:
such techniques will preferentially recover the clusters they are
designed to match, but those not fitting the model are less likely to
be identified. Many matched filter approaches also are based on
uniform background galaxy distributions, and experience a degraded
performance \citep{2002AJ....123...20K} under more realistic
backgrounds.

Our cluster detection philosophy is designed to be distinct from, but
entirely complementary to the variety of matched filter algorithms
available. This study relaxes theoretically and
observationally-motivated constraints, permitting a broader
exploration of systems with projected overdensities. Specifically, we
do not assume cluster red sequences occupy a particular position in
colour-magnitude space, nor do we stipulate preferred distributions
for the projected position of cluster members on the sky. Through this
approach we hope to provide both an independent catalogue of clusters
and a means to refine our understanding of characteristic cluster
properties. The lack of selection criteria in our algorithm permits a
double-check of the detections, since we can ask if the identified
system conforms to our expectations. As we shall later demonstrate
(see \S\ref{comparisons} and Figure \ref{MGB_J234105+00180.3}), the
prescription presented here may lead to improved recovery of certain
systems and better agreement with X-ray cluster data. Moreover,
because our proposed technique makes only two assumptions about
cluster properties, it is sensitive to a wide range of clusters,
including aspherical/asymmetric systems in the process of merging
\citep{2006ApJ...648L.109C} and fossil groups
\citep{2010A&A...514A..60S} with luminosity functions unlike a
\citet{1976ApJ...203..297S} function.

In this paper, we present our detection prescription, which involves a
blind scan of colour-magnitude space (to locate cluster sequences) and
a Voronoi tessellation technique (to estimate the galaxy surface
density distribution). Requiring only two bands to detect spectral
breaks, our approach provides a very efficient method of detecting
clusters in wide-area CCD imaging of the sky. Whilst algorithms have
in the past used Voronoi tessellations to find clusters, previous
attempts either do not exploit the red sequence or instead use
photometric redshift distribution functions that rely sensitively on
the absolute calibration and number of photometric bands
\citep{2009MNRAS.395.1845V,2011ApJ...727...45S}. In this paper we
describe the algorithm and apply it to a 7 square degree sample of
SDSS Stripe 82 data. A companion paper (GMB11) presents the full
Stripe 82 catalogue covering the full 270 square degrees.

\begin{figure*}
\centerline{\includegraphics[width=1.00\textwidth]{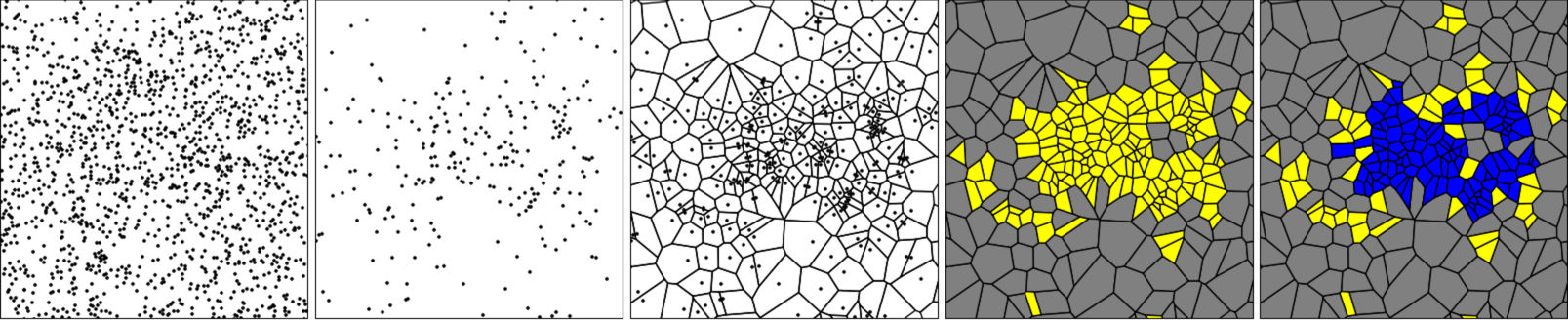}}
\caption{A depiction of the \orca detector applied to a 9'x9' cut-out
  region of Stripe 82. Starting with all galaxies in the box ({\it
    first} panel), a photometric selection (\S\ref{cuts}) isolates
  galaxies within a specific redshift range ({\it second} panel); any
  clusters in this field will be evident as surface overdensities. In
  the {\it third} panel, we compute the Voronoi diagram
  (\S\ref{voronoi}) of the distribution to estimate the surface
  density of remaining galaxies. These are separated into overdense
  (yellow) and underdense (grey) cells in {\it panel} four, according
  to how likely they are to belong to a random distribution
  (\S\ref{voronoi}). In the {\it final} panel, we use a
  Friends-Of-Friends percolation algorithm (\S\ref{fof}) to connect
  overdense cells until the density of the whole system falls below a
  density threshold. Galaxies in the blue cells become members of a
  cluster if there are at least \nmin linked members.
\label{orca_overview}
}
\end{figure*}

The outline of this paper is as follows. In section \ref{data} we define the
data used for the cluster search in the SDSS and mock catalogues. Section
\ref{method} describes the algorithm step-by-step. Section \ref{stripe82}
describes the application and testing of the algorithm using real astronomical
data, followed by a brief comparison with existing cluster catalogues in
section \ref{comparisons}. We describe the detection of mock clusters in
simulated data in section \ref{mocks}, followed by performance tests on the
simulated catalogues. In section \ref{summary} we summarise our findings.

Throughout, we assume a \lcdm cosmology with $\Omega_{\rm m}$ = 0.3,
$\Omega_{\Lambda}$ = 0.7, $H_0$ = 70 km s$^{-1}$ Mpc$^{-1}$ and $h = H_0/100$
km s$^{-1}$ Mpc$^{-1}$. For SDSS data we use the Sloan photometric system
\citep{1998AJ....116.3040G} and ``model'' magnitudes.

\section{Data}
\label{data}
\subsection{SDSS Stripe 82}
We extract Sloan Digital Sky Survey Data Release 7 {\it griz}
photometry for all sources with extinction-corrected
\citep{1998ApJ...500..525S} {\it r}-band model magnitudes $r\leq 24$
in the deep coadd stripe centred on the celestial equator (``Stripe
82'') from the SDSS Catalog Archive Server
(CAS\footnote{http://casjobs.sdss.org}). To minimise stellar
contamination, we select only galaxies where the offset between the
{\it r}-band PSF and model magnitudes satisfies $|r_{\rm PSF}-r_{\rm
model}|>0.05$. We exclude bright ($r_{\rm model}<14$) galaxies and
spurious sources such as overly de-blended galaxies and fragmented
stellar haloes. 

Although no spectroscopic or photometric redshift estimates are used
in detections, we post-process the cluster catalogue to estimate the
redshift of each system. Cluster galaxies are assigned spectroscopic
redshifts by matching source positions in the SDSS DR7, WiggleZ DR1
\citep{2010MNRAS.401.1429D} and 2SLAQ \citep{2009yCat..73920019C}
catalogues to within 1$''$. Where spectroscopic redshift data is
unavailable, we use SDSS DR7 photometric redshifts \citep[see][and
references therein]{2009ApJS..182..543A}. To increase both the source
catalogue redshift completeness and the redshift accuracy for galaxies
with no spectra, we supplement these data with additional photometric
redshifts. We select all galaxies later identified by \orca in the
GMB11 Stripe 82 catalogue and estimate their redshifts using the {\tt
hyperz} code\footnote{http://webast.ast.obs-mip.fr/hyperz}
\citep{2000A&A...363..476B} with {\it ugriz} model magnitudes and
errors. The SDSS Stripe 82 input catalogue contains 11,358,087
galaxies with Galactic extinction corrected
\citep{1998ApJ...500..525S} {\it griz} model magnitudes, over
$-50^{\circ}<\alpha<59^{\circ}$ and $\delta=\pm1.25^{\circ}$. In this
study, we concentrate on a 7 square degree sub-region within this
catalogue, centred at $(\alpha,\delta)=(355.52^{\circ},0^{\circ})$
comprising 291,389 galaxies (magnitude cuts applied to these galaxies
for cluster detection are discussed in
\S\ref{filter_parameters}). This sample, covering the same area
as the mock survey described below, was considered a large enough
observational dataset with which to test the algorithm. GMB11
describe findings from the \orca catalogue based on the full 270
square degree dataset.

\subsection{Mock Pan-STARRS Medium Deep Survey catalogue}
\citet{2009MNRAS.395.1185C} discuss the assembly of a light cone from
the Millennium Simulation \citep{2005Natur.435..629S} with a
3$^{\circ}$ opening angle, equivalent to a single pointing of the
Pan-STARRS Telescope 1 (PS-1), and the area of a single MDS tile. The
Millennium Simulation provides the \lcdm architecture into which
galaxies are populated using the \citet{2006MNRAS.370..645B} semi
analytic {\tt GALFORM} model \citep{2000MNRAS.319..168C}. This creates
a dataset with PS-1 {\it grizy} photometry for 2,346,468 galaxies down
to a magnitude limit of $r<27.5$ (equivalent to the expected $5\sigma$
depth for the PS-1 MDS) and a median redshift of $z=1.05$. The
similarity of the PS1 bands to the SDSS photometric system allows us
to apply the same magnitude limits as those set for the Stripe 82 data
(\S\ref{filter_parameters}).

\section{The method}
\label{method}
In this section we first outline, and then detail the main components
of the \orca cluster detector.

\subsection{Algorithm Outline}
\label{method_summary}
Here we describe the main steps of the \orca algorithm. With
photometry in several bands, we calculate galaxy colours in
consecutive ($g-r$, $r-i$, etc.) band pairs.

\begin{enumerate}

\item[1] We define a simple photometric selection using the colours and
magnitudes of the sample. This selection could be simple, for example a narrow
slice(s) in colour-magnitude space(s), or a more complex selection function.
This selection function can be modified in successive applications of the
algorithm to blindly scan the full photometric space, and thus isolate
red-sequences across a range of redshifts
\citep{2000AJ....120.2148G,2005ApJS..157....1G}.

\item[2] In each pass of the algorithm, we apply the photometric
  selection to the catalogue, thus greatly restricting the total
  number of galaxies under consideration. In the case of using two
  colours concurrently, this can be a very effective means of reducing
  fore- and background contamination of a putative cluster
  characterised by some red-sequence.

\item[3] After the selection, we calculate the Voronoi diagram of the
  projected distribution of galaxies on the sky. The inverse of the
  area of each convex hull surrounding each galaxy can be used as an
  estimate of the local surface density.

\item[4] Galaxies residing in dense cells (satisfying some threshold
  criteria) can be connected together into conglomerations. If enough
  galaxies are joined together in this way, we define a cluster.

\item[5] In the blind scan, successive photometric cuts may select the
  same structures (since the adjustment of the selection is by design
  less than the typical width of a red-sequence). Multiple detections
  of the same structure are identified and reduced to a single
  detection (we discuss how this was implemented in \S\ref{merging}).

\end{enumerate}

 An illustrative overview of the above procedure can be seen in Figure
 \ref{orca_overview}.

\subsection{Photometric filtering}
\label{cuts}
In large-scale imaging surveys, groups and clusters are apparent as
overdensities in the projected distribution of galaxies. Cluster
detection methods reliant only on determining the projected galaxy
density distribution are often plagued by two problems: (i) projection
effects contaminating clusters with unassociated foreground and
background galaxies (ii) the inclusion of spurious cluster detections
arising from noisy data or chance projected overdensities.
 
\begin{figure}
\centerline{\includegraphics[width=0.5\textwidth]{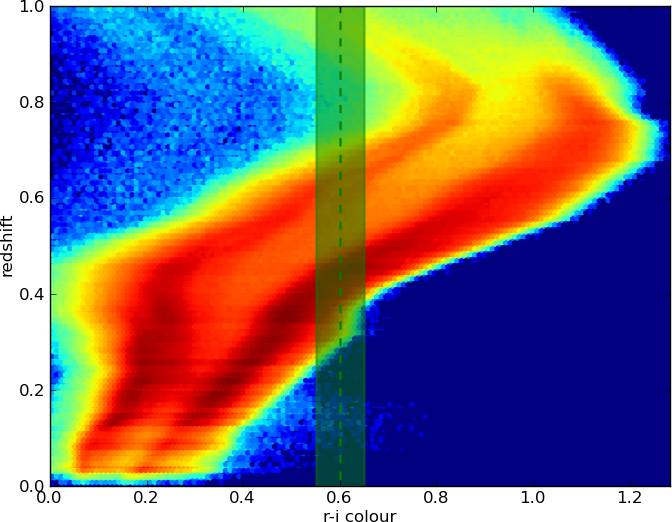}}
\caption{The redshift evolution of the observed-frame {\it r-i} colour
from a sample of mock galaxies. The colours indicate the density of
galaxies at each point, with red being the highest. We are able to
exploit this observed relation to isolate cluster galaxies within a
specific redshift range by using a selection (such as the
shaded strip in this Figure) to select galaxies from a narrow colour
range.
\label{ri_z}
}
\end{figure}

To mitigate these problems, the contrast of genuine clusters can be
enhanced by applying a photometric selection filter in
colour-magnitude space, to isolate the red-sequence ridge-line. We
parametrise our selection as a slice in colour-magnitude space,
defined by a colour-magnitude normalisation ($\cm$, the colour at
twentieth magnitude), slope $\beta(\cm)$ and width $\sigma(\cm)$. The
expected evolution of red sequence colours is constrained from simple
stellar evolution models, meaning scans over an appropriate set of
photometric selection filters allows the isolation of clusters over a
slew of redshifts. Figure \ref{ri_z} shows the redshift evolution of
galaxy colours in a sample of mock galaxies from Merson et al. (2011,
in preparation) and shows an additional advantage in using such
filters. The two tracks visibly demonstrate the bimodality in galaxy
colour that manifests itself as the ``red sequence'' \citep[lower
  track;][]{1992MNRAS.254..601B} and ``blue cloud'' (upper track). By
selecting galaxies within specific colour range $\Delta c$ (as denoted
by the green region in the Figure), one may isolate red sequence
cluster galaxies within the redshift range $\Delta z$. Contaminants in
this selection are bluer galaxies from higher redshifts. By
simultaneously selecting galaxies from two photometric selections in
different colours, one can eliminate degeneracies between colour
tracks. We discuss this further in the following section.

\begin{figure*}
\centerline{\includegraphics[width=1.00\textwidth]{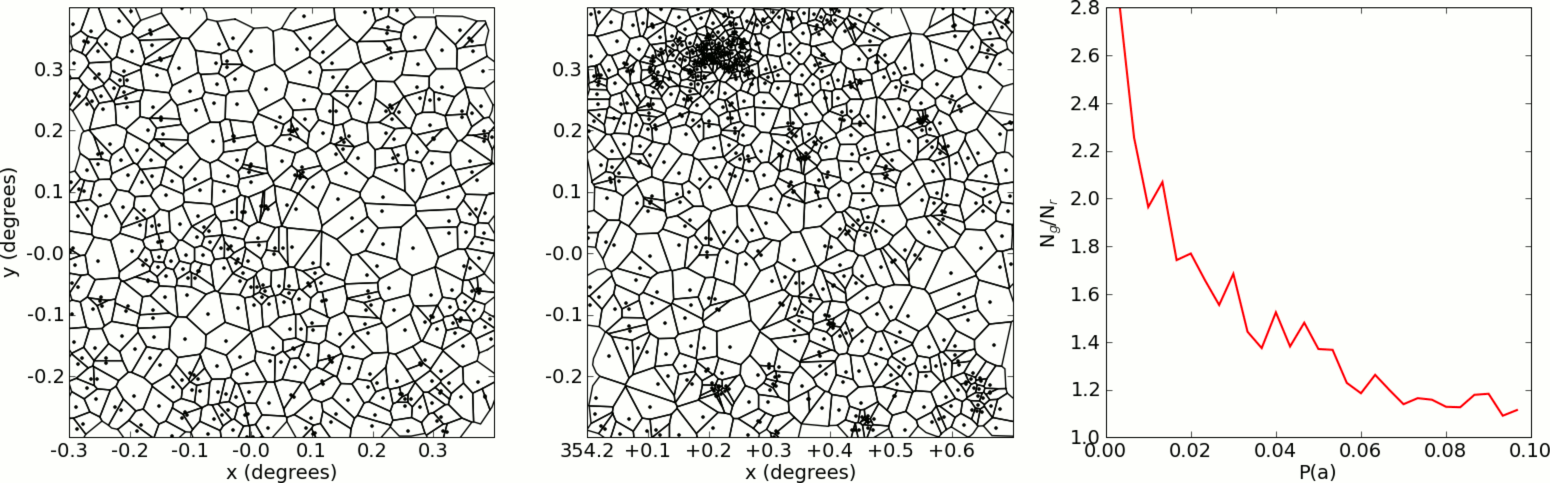}}
\caption{An illustration of the Voronoi technique described in
  \S\ref{voronoi}. The {\it (left)} panel is the Voronoi diagram of a
  random distribution of points. The {\it (middle)} panel is the
  equivalent diagram for galaxies in a field with the same mean
  density as the random field. The {\it (right)} panel shows the ratio
  of galaxy cell counts to random cell counts for a range of values of
  the integral distribution of cell areas \citep[Equation \ref{e1}
    from][]{1966ZA.....64..433K}. There is a notable excess fraction
  of galaxy cells relative to random cells at low values of P(a),
  permitting the use of a threshold to separate clustered galaxies
  from field galaxies.
  \label{vt_panel}
}
\end{figure*}

  The algorithm allows $\beta(\cm)$ and $\sigma(\cm)$ to adopt any
  values as the detector scans through colour-magnitude space. The
  simple prescription we adopt is that of a fixed slope and width with
  normalisation. Although the observed-frame sequence slope is known
  to evolve with redshift
  \citep{1998ApJ...501..571G,1998ApJ...492..461S,2009MNRAS.394.2098S},
  our choice of photometric selection width encompasses a range of
  sequence gradients large enough to account for evolution as the
  algorithm searches to deeper redshifts. Analysis of mock clusters
  from the Millennium Simulation suggests this approach probes at
  least 2.5(1.5) magnitudes fainter(brighter) than the observed
  characteristic galaxy flux at the redshifts clusters are detected in
  this study. With measurements from a large \orca cluster catalogue,
  future refinements to the algorithm may include a description of how
  the sequence slope varies with normalisation $\cm$. The values
  adopted for $\beta$ and $\sigma$ are discussed in
  \S\ref{parameters}.

We scan through colour-magnitude space in a colour $C_{\rm A}$ from
blue to red, placing down a series of M photometric selection filters
$f(C_{\rm A_{1}}),f(C_{\rm A_{2}})...f(C_{\rm A_{\rm M}})$ by increasing the
normalisation $\cm$ in small increments $dc$. The size of this
increment, set in \S\ref{filter_parameters}, allows adjacent filters
to overlap, ensuring clusters close to the boundary of a filter are
well sampled. Because each photometric selection isolates cluster
galaxies (where they exist) from a specific redshift range, the
detector can identify multiple clusters in the same line of sight. We
determine the sensitivity of the algorithm to projection in
\S\ref{projection}.

\subsection{Dual-colour photometric filtering}
\label{sub_filter}
Although only one colour is necessary to detect clusters, Figure
\ref{ri_z} notes the colour-redshift degeneracy apparent in attempting
to isolate a redshift regime from a single colour selection. One can
break the degeneracy and further reduce the field galaxy contamination
by identifying the colour range cluster members have in a second
colour $C_{\rm B}$, and subsequently applying a series of joint
photometric filters in both $C_{\rm A}$ and $C_{\rm B}$. To establish
the $C_{\rm B}$ colour range to scan, we take all cluster members from
the preliminary detection ($C_{\rm A}$ only), de-trend their sequence
slopes and fit a Gaussian to the colour distribution. The $C_{\rm B}$
colour range $\Delta C_{\rm B}$ is taken to be $\pm 1\sigma$ from the
Gaussian mean.

If the Gaussian fit is poor, detection of a clear sequence in both
$C_{\rm B}$ and $C_{\rm A}$ is less likely. In this case $\Delta
C_{\rm B}$ is simply $\pm 1\sigma$ from the median of the $C_{\rm B}$
colour distribution. The algorithm then scans over this second colour
range and attempts to detect the cluster in both colours.

A filter pair in $C_{\rm A}$ and $C_{\rm B}$ (hereafter \{$C_{\rm
  A}$,$C_{\rm B}$\}) requires a detectable sequence in both colours,
and amplifies the cluster signal by eliminating field galaxies in the
$C_{\rm A}$ filter that fail to appear within the $C_{\rm B}$
filter. Any cluster in the final catalogue detected in $C_{\rm A}$
must therefore also have been detected in $C_{\rm B}$. This improves
the robustness of the algorithm and the reduction of contaminants from
spurious detections. Because sub-filters overlap in $C_{\rm B}$
colour-magnitude space, the same cluster may be detected in multiple
filters. We apply the prescription described in \S\ref{merging} to
identify and merge clusters that have been detected in more than one
filter. The number of selection filters used to sample any colour
range depends on the sampling interval $dc$ set in
\S\ref{filter_parameters}.

\subsection{Identifying overdensities with the Voronoi tessellation}
\label{voronoi}
After increasing a cluster's detectability by suppressing field
galaxies with photometric filters, the next step is to calculate the
local surface density of each galaxy. Galaxies residing in common
regions of enhanced density can then be grouped together into
clusters. To quantify the surface density field, we divide the
galaxies into Voronoi cells using {\tt
qhull}\footnote{http://www.qhull.org}
\citep{Barber96thequickhull}. The Voronoi diagram is a tessellation of
convex hulls, or cells, with each galaxy occupying only one cell. All
positions inside a given cell are closer to the cell's nucleus (the
galaxy) than any other. Unlike many other detection techniques, the
Voronoi Tessellation \citep[for VT cluster detection,
see][]{1993PhRvE..47..704E,2001A&A...368..776R} does not smooth the
data, is robust to cluster ellipticity \citep[][]{1991MNRAS.249..662P}
and can be applied to a variety of survey geometries. VTs do not
suffer from spurious detections around survey boundaries and edges,
and are thus well suited to analysing astronomical data with localised
camera defects, excised bright stars and other sources of
incompleteness. The {\it left} and {\it middle} panels of Figure
\ref{vt_panel} respectively show the Voronoi diagrams for a random
point distribution and galaxies with identical mean densities
$\bar{\Sigma}$. Galaxies in more concentrated regions tend to have
smaller cells.

We define the reciprocal of the galaxy cell area ($\rm a_{\rm g}$) as an
estimate of the galaxy's local surface density
$\hat{\Sigma}_{\rm g}$. Searching for connected regions of high density
identifies statistically significant structures. To determine if a
galaxy resides in a high density region of the survey, we evaluate the
statistical significance of finding a cell of area $\rm a_{\rm g}$ in a random
field with mean cell area $\bar{\rm a}_{\rm R}$. We use the
\citet{1966ZA.....64..433K} cumulative function for a Poissonian distribution
of points: 

\begin{figure*}
\centerline{\includegraphics[width=1.00\textwidth]{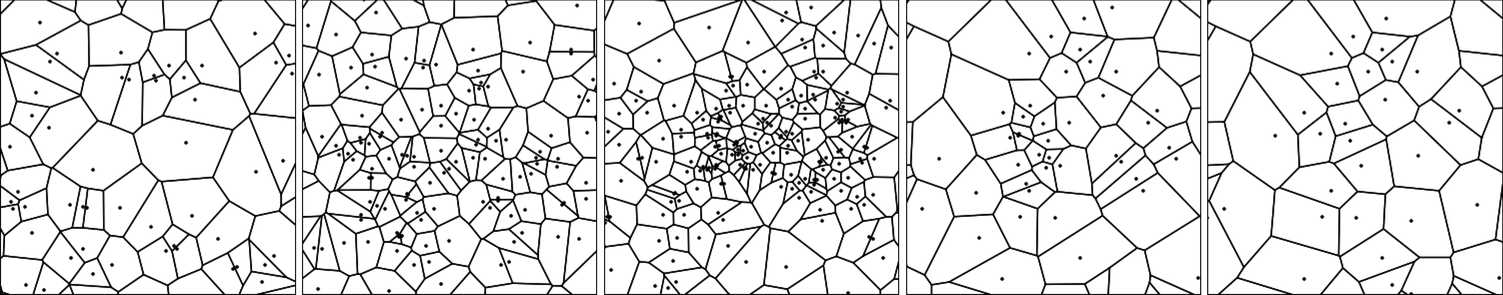}}
\caption{A sequence of Voronoi diagrams generated from galaxies in the
same area of sky, but selected from different photometric filters. A
cluster signal is apparent for some filters, but is not apparent in
others. This demonstrates the power of colour selection in isolating
galaxies at specific redshifts. In cases where a cluster may be
detected in more than one filter (such as the borderline detection in
the second panel), the algorithm must decide which cluster to
select. This aspect of the detector is discussed in \S\ref{merging}.
\label{abell_voronoi}
}
\end{figure*}

\begin{equation}
\label{e1}
P(a) = \int_0^a {\rm d}p = 1 -  e^{-4a}\left(\frac{32a^3}{3} + 8a^2 + 4a + 1 \right)
\end{equation}

where $a=(\rm a_{\rm g}/\bar{\rm a}_{\rm R})$. The {\it right} panel of
Figure \ref{vt_panel} shows the distribution P(a) for cells in an
example galaxy field relative to a Poisson distribution of the same
field size and number of points. Candidate cluster galaxies residing
in overdense regions can be selected by cell areas statistically
unlikely to arise in a random distribution. An excess of galaxy cells
is apparent for low P(a) compared to the random distribution. We
identify all galaxies with $P(\rm a_{\rm g})<\probthresh$ in order to select a
population of clustered galaxies. The choice of overdensity
probability threshold is discussed in \S\ref{vt_parameters}.

\subsection{Connecting overdense regions to form clusters}
\label{fof}
Remaining galaxies belonging only to overdense cells are now grouped
together to form clusters. We achieve this by applying a
Friends-Of-Friends algorithm to these cells. Rather than a distance
criterion, we define a ``friend'' as an adjacent Voronoi cell sharing
at least one vertex. Potential clusters are seeded by ordering the
cells with decreasing density, iterating through and connecting
adjacent cells. These overdense regions grow by percolation until
either no more adjacent overdense cells remain, or the mean cell
density of the putative cluster:

\begin{equation}
\label{e2}
\bar{\Sigma}_{\rm cells}=\rm N_{\rm gal}\sum_{i=1}^{\rm N_{\rm gal}}\frac{1}{a_{i}}<\Sigma_{crit}
\end{equation}

Groups of connected galaxies are classified as clusters if they have
$\rm N_{\rm gal}\geq~$\nmin. The choice of the critical density
threshold $\Sigma_{\rm crit}$ and \nmin~algorithm parameters is
discussed in \S\ref{filter_parameters}.

\subsection{Producing a cluster catalogue}
\label{merging}
In \S\ref{cuts} and \S\ref{sub_filter} we noted that adjacent
photometric filters applied to the input catalogue overlap in
colour-magnitude space. With this sampling strategy, the same cluster
could be detected in multiple filters. Figure \ref{abell_voronoi}
shows a sequence of Voronoi tessellations applied to the same area of
sky under photometric filters sensitive to different redshift
ranges. Because colour scans sample the colour range of a red-sequence
at a specific redshift, the cluster will be detected in multiple scans
(with a peak contrast where the selection is most effective). In cases
of clusters detected multiple times in different photometric filters,
the ``best'' cluster is identified and added to the final cluster
catalogue.

\begin{figure*}
\centerline{\includegraphics[width=1.00\textwidth]{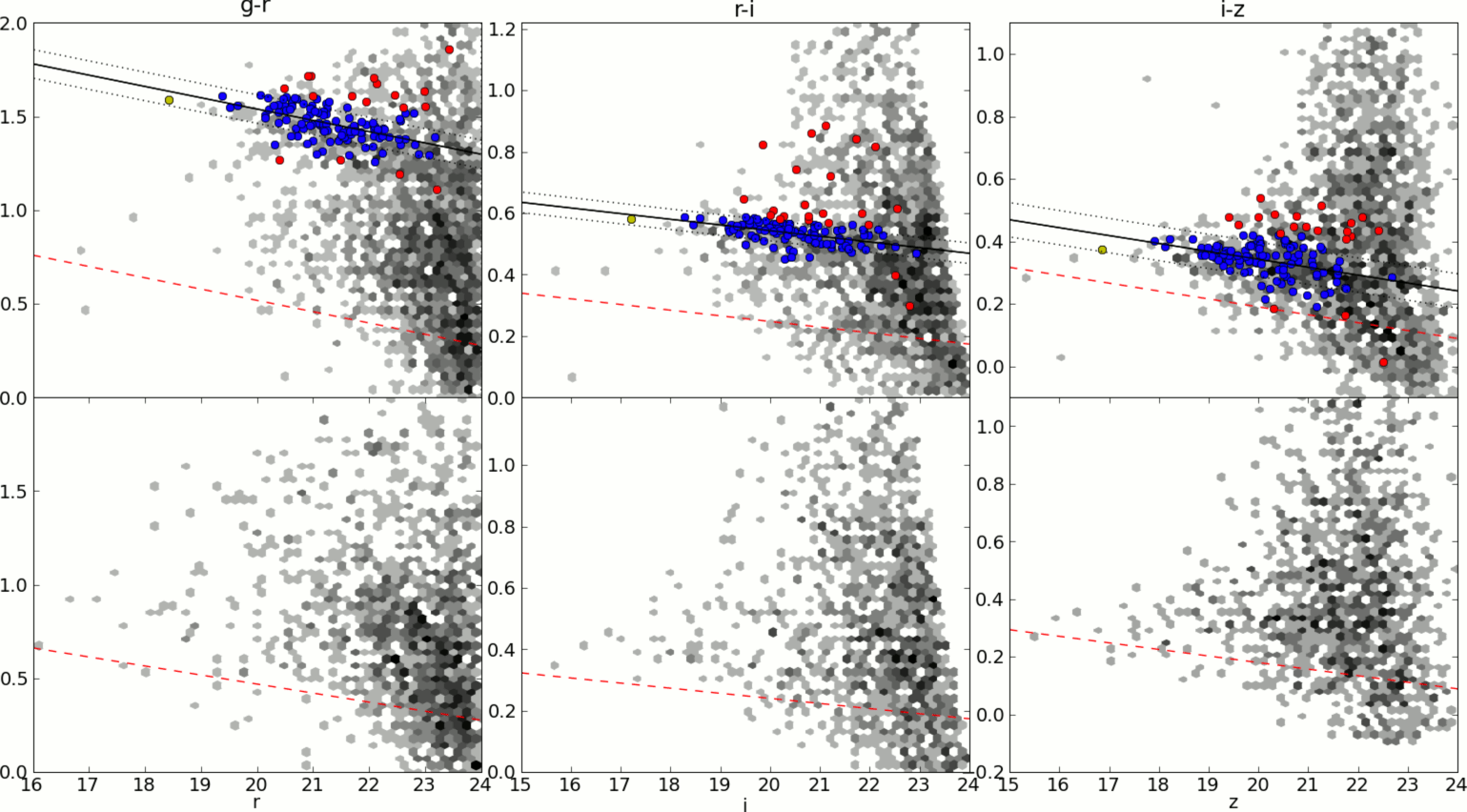}}
\caption{({\it Top}) Colour-magnitude diagrams for the 126 Abell 2631
  members selected in this study. The yellow dot notes the position of
  the cluster {\it r}-band brightest cluster galaxy. The black lines
  denote photometric selection filter fits to the data and indicate
  the slope ($\beta$), normalisation (solid, $\cm$) and width (dotted,
  $\sigma$). The identified members are split into those inside (blue)
  and outside (red) the 3-sigma cut used to estimate the filter
  width. Grey data indicate all galaxies that were not identified as
  members of the cluster out to a radius of 7-arcminutes from the
  cluster centre. The red dashed line in the {\it g-r} colour
  indicates the blue limit imposed by the Virgo cluster, and the
  equivalent lines in {\it r-i} and {\it i-z} denote the lowest $\cm$
  identified from cluster sequences in our search of the 7 square
  degree Stripe 82 survey. ({\it Bottom}) The colour-magnitude
  diagrams for galaxies in a region of the same area located in a
  field environment.
\label{abell_cmr}
}
\end{figure*} 

For two candidates to be considered detections of the same system,
they must have sufficiently similar spatial positions, red sequence
fits and cluster members. We quantify the similarity in cluster
sequences using linear fits to the colour-magnitude relation for the
galaxies in each cluster detected. Sequence slopes can in principle
adopt any value permitted by the width of the photometric filter
(defined here as $\sigma_{f}$) it was selected in. We quantify the
similarity between two sequences with the following criteria:

\begin{enumerate}
\item[-~] {\it Sequence match 1} ($\Delta \rm S_{1}$): True if the sequence
separation is $<0.5\sigma_{f}$ in colour for at least 25\% of the magnitude
range $m_{\rm BCG}\leq m \leq m_{\rm BCG}+5$.

\item[-~] {\it Sequence match 2} ($\Delta \rm S_{2}$): True if the sequence
separation is $<\sigma_{f}$ in colour difference for at least 50\% of the
range described in $\Delta \rm S_{1}$.

\item[-~] {\it Sequence match 3} ($\Delta \rm S_{3}$): True if the colour
difference at 20$^{th}$ magnitude, ($\Delta\cm$) between the two sequences is
$<\sigma_{f}$.

\item[-~] {\it Sequence match 4} ($\Delta \rm S_{4}$): True if the clusters were
detected in adjacent (overlapping) filters. 
\end{enumerate}

To define the similarity in cluster membership, spatial position and
extent, we describe the {\it common-galaxy} fraction and {\it
projection extent} for two clusters, $\rm CL_{1}$ and $\rm CL_{2}$:

\begin{enumerate}
\item[-~] {\it Common galaxies} ($\rm cg_{\rm 1,2}$): the fraction of galaxies in
$\rm CL_{1}$ that also belong to $\rm CL_{2}$. Similarly, $\rm cg_{\rm 2,1}$ is the
fraction of $\rm CL_{2}$ galaxies also appearing in $\rm CL_{1}$. The ${\rm BCG}_{id}$
boolean notes when clusters share the same BCG.

\item[-~] {\it Projection extent} ($\rm pe_{\rm 1,2}$): the fraction of galaxies
in $\rm CL_{1}$ that lie within the Voronoi cell boundaries of the $\rm CL_{2}$
cluster. As with $\rm cg$, $\rm pe_{\rm 2,1}$ is the case for $\rm CL_{2}$.
\end{enumerate}

With these measures, five tests of ``cluster similarity'' were devised
(Table \ref{constraint_table}). A pair of clusters must pass {\it at
least one} to be considered detections of the same system. Each of
these tests account both for the spatial and colour characteristics of
the clusters. Because no merging can proceed purely by colour
similarity or spatial coincidence, this ensures the separation of
associated but distinct systems, and clusters in projection. We
balance these requirements with the need to prevent multiple instances
of the same cluster appearing in the final catalogue. Where matches
between two clusters exist, the thresholds in Table
\ref{constraint_table} make it likely the two systems will be
merged.

To define the ``best'' cluster from a list of candidates, we pick out
the system with the largest {\it reduced flux} - the total flux (in
the detected band) of all but the three brightest cluster
members. This prevents the selection of a cluster including one or two
bright galaxies that may not be genuine members, but also makes the
choice of best cluster largely independent of the BCG. Once the
``best'' cluster is selected, the remaining candidates are discarded
from the catalogue. However, to each cluster selected in this way, we
attach a record of the candidate cluster galaxies that were not
selected, forming an auxiliary catalogue of {\it associate cluster
members}.  In this way, we can keep track of galaxies the detector
considered as members but did not include in the cluster. The degree
of oversampling in colour space and hence number of multiple
detections depends on the sampling interval $dc$, relative to the
width $\sigma(\cm)$ of the filter. We set both of these parameters
in \S\ref{filter_parameters}.

\begin{table}
\begin{center}
\caption{The set 
of conditions used to consider whether two clusters are multiple
  detections of the same system. If any one of these conditions are
  satisfied, the algorithm picks the ``best'' cluster of the two.}
\begin{tabular}{|l|l}
 \hline
 \# & Constraint \\
 \hline
 1 & ($\rm cg_{1,2}$ OR $\rm cg_{2,1}$) $\geq 0.5$\\
 2 & ($\rm pe_{1,2}$ OR $\rm pe_{2,1}$) $> 0$ AND $\Delta \rm S_{1}$\\
 3 & $\rm BCG_{id}$ AND $\Delta \rm S_{2}$\\
 4 & ($\rm pe_{1,2}$ OR $\rm pe_{2,1}$) $\geq 0.8$ AND $\Delta \rm S_{3}$\\
 5 & ($\rm pe_{1,2}$ OR $\rm pe_{2,1}$) $\geq 0.8$ AND $\Delta \rm S_{4}$\\
 \hline
\end{tabular}
\label{constraint_table}
\\
\end{center}
\end{table}

\subsection{Algorithm parameters}
\label{parameters}
 This section defines the values adopted for the algorithm parameters
 described in \S\ref{cuts}-\S\ref{fof}.
\subsubsection{Photometric filtering}
\label{filter_parameters}
 In both mock and real datasets, we limit our search for clusters to
 three colours: {\it g-r}, {\it r-i} and {\it i-z}. These are used to
 form joint selection filters combining two colours: \{{\it g-r}, {\it
   r-i}\} and \{{\it r-i}, {\it i-z}\}. 

 Each photometric filter is described by a colour normalisation $\cm$,
 slope $\beta(\cm)$ and width $\sigma(\cm)$. For this study
 and that of GMB11 we demonstrate the detector with an
 unchanging filter slope and width. In order to set $\beta$ and
 $\sigma$ for each colour, 126 members of Abell 2631
 (\citealt{1989ApJS...70....1A}) are visually identified in an {\it
   i}, {\it r} and {\it g} composite Stripe 82 image. At redshift
 $z=0.278$ \citep{2000ApJS..129..435B}, this system is the richest
 Abell cluster in Stripe 82 and shows evidence of a clear sequence in
 all three colours used in this study.

\begin{figure}
\centerline{\includegraphics[width=0.5\textwidth]{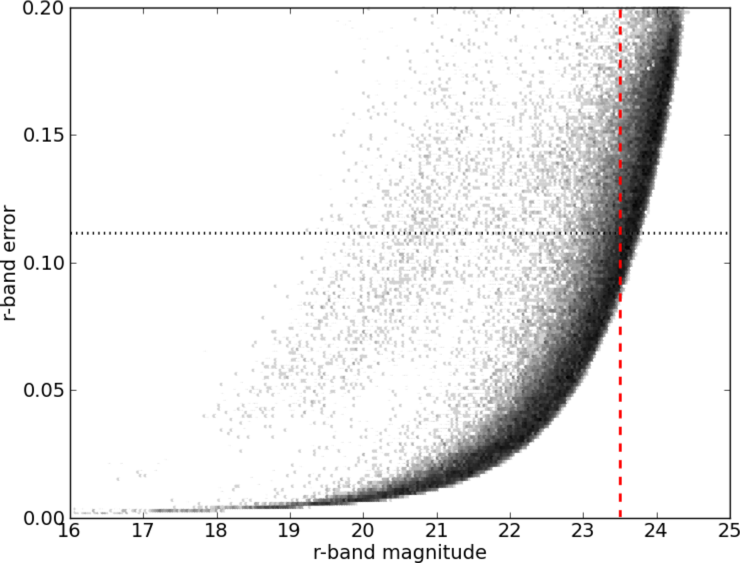}}
\caption{The SDSS model {\it r}-band photometric error in a sample of
  100,000 Stripe 82 galaxies. These data are used to set a magnitude
  limit where at least 50\% (0.68$\sigma$, black horizontal dotted
  line) of the faintest galaxies remain in a colour slice of width
  $\sigma_{f}=0.152$. Whilst the data suggest a limit of $r\leq23.8$,
  we opt for a slightly more conservative $r\leq23.5$ limit (red
  vertical dashed line).
\label{phot_error}
  }
\end{figure}

A linear fit to the colour-magnitude sequence was applied to determine
$\beta$ for each colour. The filter widths were set using a method
akin to that described in \citet{1998ApJ...501..571G}; we first remove
the slope in each sequence and then exclude $3\sigma$
outliers. Starting at the line fitted to the cluster sequence, we
increase the width in equal amounts above and below this line until we
enclose 90\% of the remaining members. We define this as the filter
width $\sigma$ for that colour.

Figure \ref{abell_cmr} shows the colour-magnitude sequence of the
identified members in the three colours ({\it top}) compared to a
field of the same area with no cluster present ({\it bottom}). Blue
(red) points identify members that were inside (outside) the $3\sigma$
cut used to identify outliers. Grey data correspond to galaxies that
were within $7'$ of the cluster centre and not picked as cluster
members. Table \ref{filter_table} lists the fitted filter
parameters for each colour (corresponding to the black lines in Figure
\ref{abell_cmr}) in addition to the colour range and number of
filters used in our cluster search. Following our decision in
\S\ref{cuts} to use a fixed slope, we adopt the largest filter
width ($\sigma_{f}$, 0.152) for all colours, and use this to define
the input galaxy magnitude limit for each band. Magnitude limits are
applied to reduce the number of input galaxies with high levels of
photometric uncertainty. We set these as the faintest magnitude where
the photometric uncertainties fall below $0.68\sigma_{f}$

We set limits for each band based on a sample of 100,000 galaxies from
{\it Stripe 82}. Figure \ref{phot_error} shows the galaxy photometric
error distribution for the {\it r}-band, and from this we set a
magnitude limit of $r\leq23.5$. This is slightly more conservative
than the limit implied by the error distribution ($r\leq23.8$) because
we aim to include only sources with good photometry. The magnitude
limits applied are 24.0, 23.5, 23.3, 21.6 in the {\it g}, {\it r},
{\it i} and {\it z} bands respectively, resulting in a source
catalogue of 69,797 galaxies. With the added depth from Stripe 82
photometry, these limits permit an exploration of the red sequence to
at least 2.5, 3 and 1.5 magnitudes fainter than M$_{\star}$
respectively for the {\it r}, {\it i} and {\it z} bands. As part of
the algorithm design, we considered multiple searches through the data
at different flux limits. Under this prescription, higher-signal
cluster sequences would be selected when re-detections of the same
system were merged. In tests with the mock lightcone data analysed in
\S\ref{mocks}, we found no significant advantage in this
implementation, and instead kept our magnitude limits fixed.

The bluest filter pair we employ is {\it g-r}. To prevent the detection of
spurious systems bluer than the $z=0$ red-sequence in this colour we determine
a blue limit by extrapolating the colour-magnitude relation (CMR) for Coma
\citep{2009MNRAS.392.1265S} and Virgo \citep{2008AJ....135.1837R} to {\it
r}$=20$. The $\cm$ normalisation for Coma (Virgo) was estimated as 0.6
(0.47); we use the latter as the bluest filter possible in the {\it
g-r} colour. We do not apply similar limits to the other colours, but
the normalisation below which no sequences were detected in {\it r-i}
and {\it i-z} is described in \S\ref{the_catalogue}. Figure
\ref{abell_cmr} shows these limits as red dashed lines.

\begin{figure*}
\centerline{\includegraphics[width=1.0\textwidth]{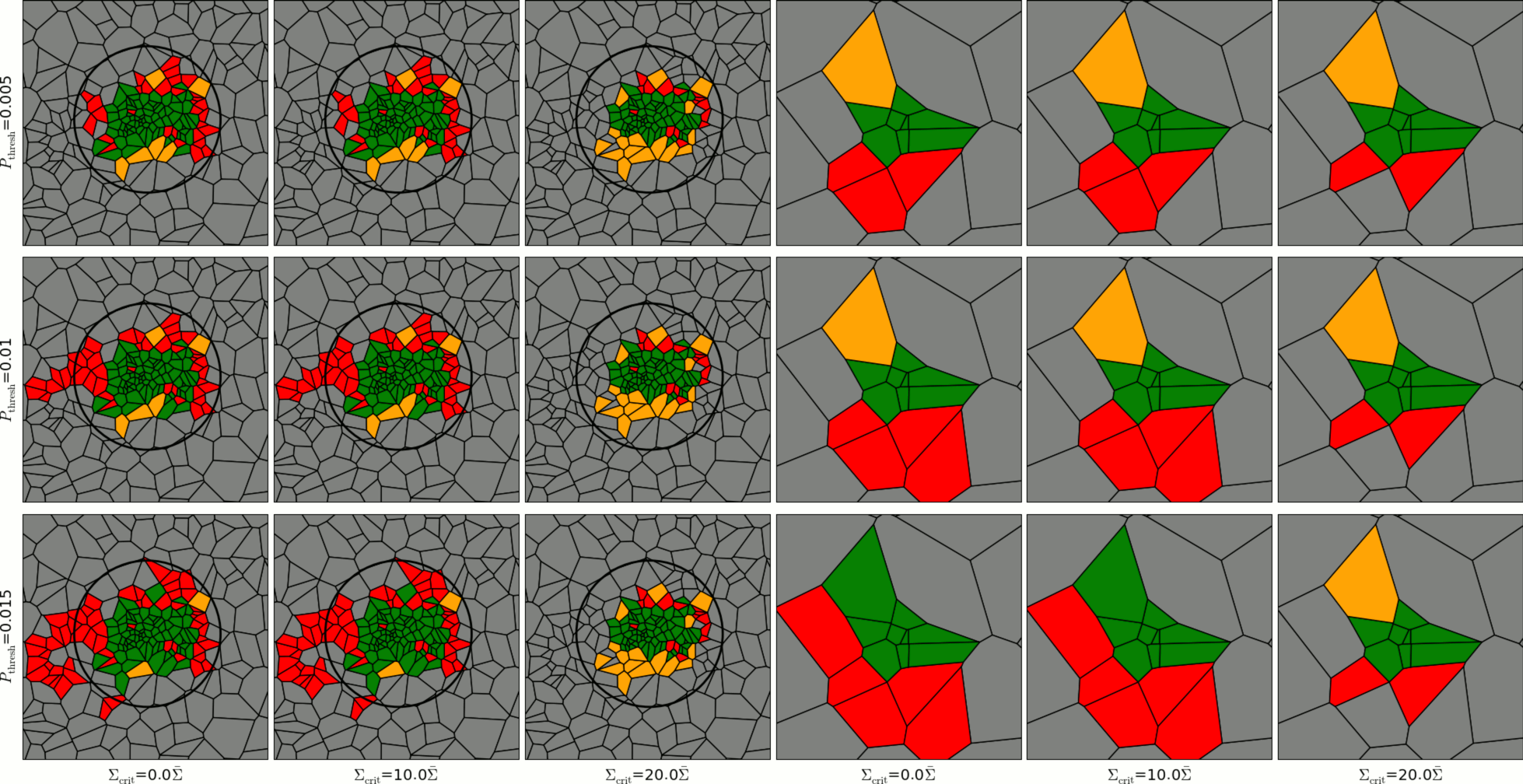}}
\caption{Effect of detection parameters on Abell 2631 ({\it left},
  box scale 13.6'$\times$13.6') and a compact group ({\it right}, box
  scale 3.5'$\times$3.5'). Colour key: Grey are cells with field
  galaxies, green are galaxies identified by the algorithm that were
  also visually identified as members. Red cells are members assigned
  to the cluster by the detector but not visually identified as
  cluster members.  Orange cells are galaxies that failed to be
  correctly identified by the algorithm as cluster galaxies, but were
  defined as such visually. The circle around Abell 2631 corresponds
  to a $1\mpc$ radius at the cluster redshift.
\label{parameter_plots}
  }
\end{figure*}

\begin{table}
\begin{center}
\caption{Filter parameters fitted from Abell 2631, the ranges searched
  and the number of filters in each colour. The blue limit in {\it
    g-r} corresponds to an extrapolation of the Virgo CMR, whilst the
  others permit a full sweep of the available data. The emboldened
  figure is the largest filter width ($\sigma_{f}$), and is adopted
  for all colours.  }
\begin{tabular}{|c|c|c|c|c} \hline
Colour & Slope ($\beta$) & Width ($\sigma$) & Range & Filters\\
\hline
{\it g-r} & $-0.048$ & $\bf{0.152}$ & $0.47-2.00$ & $39$ \\
{\it r-i} & $-0.017$ & $0.067$ & $0.00-1.22$ & $38$ \\
{\it i-z} & $-0.023$ & $0.110$ & $-0.10-1.10$ & $31$ \\
\hline
\end{tabular}
\label{filter_table}
\\
\end{center}
\end{table}

 Finally, the detection algorithm uses photometric filters that
 overlap in colour-magnitude space, preventing clusters close to
 filter edges from being poorly sampled.  A sampling interval in
 colour space of $dc=0.04$ is chosen, corresponding to an overlap of
 approximately 75\% between adjacent filters based on $\sigma_{f}$,
 the filter width.

\subsubsection{Voronoi Tessellation and connection of overdense regions}
\label{vt_parameters}
The initial identification of clusters in projected high density
regions and the subsequent percolation of their members depends
respectively on the probability threshold $\probthresh$ and the
critical density \sigmacrit. We parametrise the critical density
\sigmacrit as a scalar multiple of $\bar{\Sigma}$ such that both
detection parameters have a mean density dependence. In the {\it
left-hand} sequence of Figure \ref{parameter_plots}, we note the
effect a range of ($\probthresh$,\sigmacrit) combinations have on the
recovery of Abell 2631 within a box of scale 13.6'. By tracking the
detector's assignment of Voronoi cells to cluster and field, we
compare members visually identified to the recovery of this cluster
under different parameter combinations. The cells are colour-coded
into four groups to differentiate detected and visually identified
members. Grey cells show galaxies neither detected nor identified as
cluster members. Green cells denote detected members that were also
visually identified, orange for where the detector did not assign
cluster membership despite our classification as such from the
imaging, finally red cells are detected members not visually
identified as members. We stress the latter group in no way indicates
the purity of the cluster, as we are both incomplete and subjective in
our identification of genuine cluster members. However, this exercise
does provide a useful indication of detector performance when compared
to our visual impression of cluster membership.

The detection grids show re-detection is broadly insensitive to the
range of parameters explored. At higher probability thresholds
(increasing row number) the cluster expands to form a more extended
structure. This growth is moderated by the introduction of a minimum
cell density.  We exclude \sigmacrit$=20\bar{\Sigma}$ as it removes a
significant fraction of visually identified members on the periphery
of the cluster. The middle ground between detecting a more compact
system ($\probthresh$=0.005) and potentially increasing the interloper
fraction ($\probthresh$=0.015) suggests the balance of detection
completeness and cluster purity lies with $\probthresh$=0.01. We note
from Figure \ref{vt_panel} there are at minimum twice as many
clustered cells as unclustered at $P(a)\leq0.01$. Although
(0.01,$0\bar{\Sigma})$ and (0.01,$10\bar{\Sigma}$) appear identical in
their recovery of the cluster, we require a non-zero density
constraint to filter out spurious low amplitude systems and prevent
large clusters from percolating into giant connected structures. We
consequently adopt the parameter combination
($\probthresh$,\sigmacrit)=(0.01,10$\bar{\Sigma})$. To ensure these
parameters are not biased to the detection of high mass systems, we
use 11 members of a visually identified compact group to perform a
re-detection in the same parameter ranges. The {\it right-hand}
sequence in Figure \ref{parameter_plots}, with boxes of scale $3.5'$,
shows the recovery of this group, and indicates group scale detection
is robust to the range of parameters explored. The trade-off between
completeness and purity is similarly evident here, with
$(0.01,10\bar{\Sigma})$ remaining a good compromise between the two.

In both cases (and more generally) there is a tendency to
underestimate the total number of cluster members. This arises from an
inevitable feature of Voronoi Diagrams implying the algorithm is
unlikely to recover all cluster members. The suppression of the field
galaxy population with photometric filters causes an abrupt drop in
galaxy surface density at the cluster boundary. Because the Voronoi
cells of peripheral members have a limited number of field galaxies to
constrain their boundaries they adopt larger areas. Such cells may
then be rejected as members because their areas are inconsistent with
that population. Nevertheless, tests with mock catalogues allow us to
quantify the impact this effect has on the cluster purity, as
discussed later in \S\ref{mocks}.

 Finally, we set the minimum membership of a cluster, \nmin, to
 five galaxies.

\section{SDSS Equatorial stripe 82 cluster catalogue}
\label{stripe82}
\subsection{The catalogue}
\label{the_catalogue}
We applied the detector to a 7 square degree sample of Stripe 82,
using the limits described in \S\ref{data} and parameters
described in \S\ref{parameters}. Here we describe the general
characteristics of this catalogue, perform a series of tests on the
data and briefly compare our detections to existing optical and
X-ray-detected clusters.

After applying the magnitude limits described in
\S\ref{filter_parameters}, a source catalogue of 69,797 galaxies is
analysed by the algorithm. We find a total of 97 clusters, identifying
a total of 1293 cluster galaxies (0.5\% of the original galaxy sample)
and 813 associate cluster members (candidate cluster members that were
not selected). Of these clusters, 34\% were detected in \{{\it g-r},
{\it r-i}\} and 66\% in the \{{\it r-i}, {\it i-z}\} combinations.

Although we define a blue limit for the {\it g-r} colour-magnitude
relation ($\cm>0.47$), equivalent limits were not applied to the {\it
r-i} and {\it i-z} colours. We can however place upper bounds on the
blue limit in these colours by noting no clusters were detected below
{\it r-i}=0.24 and {\it i-z}=0.18. Such limits serve to reduce the
search time for future survey scans.

Table \ref{cat_sample} shows an extract of the cluster catalogue. This
7 square degree sample of 97 Stripe 82 clusters is available
online\footnote{http://orca.dur.ac.uk/}. Each cluster is named
according to the IAU convention, in the form MGB JHHMMSS+DDMM.m. We
detail below the main features both catalogues.

\begin{table*}
\begin{center}
\caption{A sample of the \orca 
cluster catalogue generated in this study. Full details of the columns
  can be found in \S\ref{the_catalogue}-\S\ref{r80}. The
  first column contains the cluster name based on the IAU
  convention. Columns 2 and 3 note the J2000 estimated cluster
  positions in degrees. Columns 4 and 5 describe the cluster redshift
  and source data used to calculate the redshift. Column 6 notes how
  many members were found in the cluster, and we provide estimates for
  the cluster $B_{\rm gc}$ richness and sequence scatter in Columns 7
  and 8.  The final two columns indicate the radius (in degrees)
  enclosing 80\% of the cluster members and the ratio of this value to
  the 20\% radius, a measure of cluster concentration.}

\begin{tabular}{|c|c|c|c|c|c|c|c|c|c}
\hline
Name & RA & DEC & cluster\_z & cz\_type & $\rm N_{\rm gal}$ & b\_gc & scatter & $\theta_{\rm 80}$ & C \\
\hline
MGB J234017-00030.9 & 355.06912 & -0.06455 & 0.245 & c0s0w0q0d0b0p6h2 &     6 & 19416 & 0.047 & 0.0001 & 1.700 \\
MGB J233817+00190.0 & 354.56897 & 0.33309 & 0.208 & c0s0w0q0d0b0p8h6 &     8 & 94461 & 0.038 & 0.0003 & 3.667 \\
MGB J234113-00000.4 & 355.30349 & -0.00597 & 0.166 & c0s0w0q0d0b0p6h2 &     6 & 182181 & 0.018 & 0.0003 & 1.692 \\
MGB J234400-00300.3 & 355.99952 & -0.50461 & 0.181 & c0s1w0q0d0b0p5h4 &     6 & 71831 & 0.025 & 0.0001 & 1.750 \\
MGB J234725+00190.7 & 356.85322 & 0.32867 & 0.201 & c0s0w1q0d0b0p14h14 &    14 & 10967 & 0.037 & 0.0004 & 2.545 \\
\hline
\end{tabular}
\label{cat_sample}
\end{center}
\end{table*}

\subsection{Cluster positions \& redshifts ({\tt cluster\_z}, {\tt cz\_type})}
\label{cz}

 The {\tt ra} and {\tt dec} position quoted in the catalogue is the
 algorithm estimate of the centre of each cluster, based on the
 average positions of their members.

 Although we do not use any redshift data to generate our cluster
 catalogue, we provide redshift estimates for each system detected by
 the algorithm. These redshifts are weighted towards members with
 spectroscopic data, but two sets of photometric redshift data ({\it
   hyperz} and the DR7 photometric estimate) are used to provide each
 cluster galaxy with at least one redshift estimate. From the
 catalogue of 1293 cluster galaxies, 2.6\% have spectroscopic data
 (DR7 spectroscopic redshifts, WiggleZ and 2SLAQ), 93\% have DR7
 photoz and 87\% have {\tt hyperz} estimates. The {\tt hyperz}
 estimates for cluster members were generated using only S0 and E
 SEDs, a \citet{2000ApJ...533..682C} reddening law and a two-stage
 convergence (over and above that performed by {\tt hyperz}) to the
 redshift where a range identified in coarse redshift bins is
 re-sampled with a smaller bin width. Comparing these estimates to
 available spectroscopic redshifts, the measured error dispersions are
 higher in {\it hyperz} than in the DR7 pipeline (0.029 vs 0.016).

 We calculate each cluster redshift by determining the weighted median
 redshift from the available member data. The weighting for a
 spectroscopic, DR7 photoz and {\tt hyperz} redshift is 4, 2, 1
 respectively, the higher weighting for DR7 photoz reflecting the
 smaller error dispersion mentioned above. To gauge the accuracy of
 our redshift estimate, we note the calculated redshift of Abell 2631
 is $z=0.26$, some 0.02 lower than the value determined
 by \citet{2000ApJS..129..435B}. The median cluster redshift of the
 whole catalogue is $z_{\rm med}=0.31$, and the maximum redshift is
 $z=0.57$. Approximately 25\% of the clusters have at least one member
 with a spectroscopically measured redshift.

\begin{figure*}
\centerline{\includegraphics[width=1.0\textwidth]{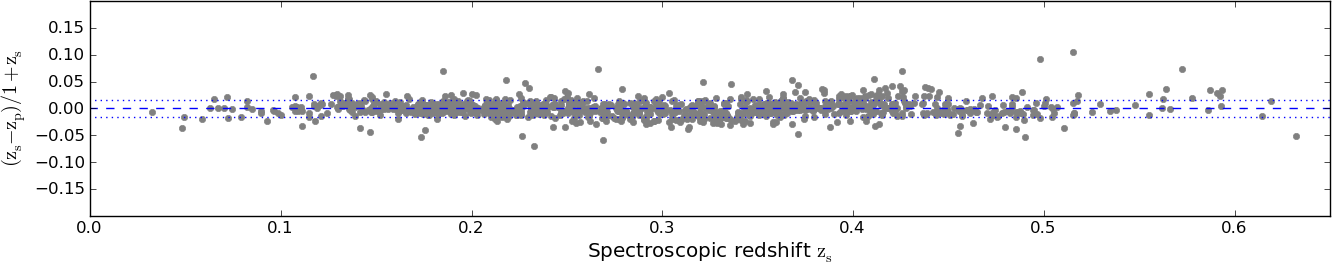}}
\caption{Comparison of photometric redshift accuracy $\delta_{\rm
    z}(z_{\rm s})=(z_{\rm s}-z_{\rm p})/1+z_{\rm s}$ for the cluster
  BCGs with spectroscopic redshifts. After outlier rejection (clipping
  galaxies with $|\delta_{\rm z}| > 3\sigma_{\delta_{\rm z}}$, or
  0.4\% of the total sample) and removing the slight systematic photoz
  error, we find a 1-$\sigma$ scatter $\sigma_{\delta_{\rm z}}$=0.0157
  (denoted by the dotted blue lines). This highlights the excellent
  redshift recovery using {\it ugriz} photometry alone. For a given
  cluster we combine both the photometric and (where available)
  spectroscopic redshifts of cluster members to derive a robust
  redshift estimate for the system as a whole.
\label{sz_acc}
  }
\end{figure*}

 Without access to spectroscopy, accurate photometric redshifts of red
 sequence cluster galaxies are good measures of cluster redshifts. We
 quantify this in Figure \ref{sz_acc} by comparing the photometric and
 spectroscopic redshifts of cluster BCGs from a sample of the full
 GMB11 Stripe 82 cluster catalogue with spectroscopic redshifts. After
 removing a small systematic trend and $3\sigma$ outliers, the
 $1\sigma$ dispersion in $(z_{\rm s}-z_{\rm p})/1+z_{\rm s}$ is 0.0157
 (increasing to 0.0163 when ignoring the systematic error). This
 suggests BCG photometric redshifts are accurate estimates of the
 cluster redshift.

The {\tt cz\_type} property is a shorthand description of the
available redshift data for each cluster, each letter defining a
measurement type, followed by the number of that type. The letters
denote data from the mo(c)k, DR7 (s)pectroscopic, (w)iggleZ, 2SLA(q),
DR7 (p)hotometric and (h)yperz datasets, where mock is of course not
used in this observational data.

\subsection{Cluster richness ({\tt b\_gc})}
\label{b_gc}
 With access to cluster redshifts we are able to calculate the $\rm B_{\rm gc}$
 optical cluster richness, a robust parameter known to correlate with
 cluster mass. We use the $\rm B_{\rm gc}$ measure described in
 \citet{1999AJ....117.1985Y}:

\begin{equation}
\label{bgc}
\rm B_{\rm gc}= {\rho_{\rm bg} \rm D({\it z}_{\rm cl})^{\gamma-3} {\rm A_{\rm gc}}\over{I_\gamma\Phi(M_{3},M_{3}+3,{\it z}_{cl})}}
\end{equation}

 where $\rho_{\rm bg}$ is the background surface density of all source
 catalogue galaxies (irrespective of their colour) inside a $0.5\mpc$
 radius with luminosities between the third brightest cluster galaxy
 ($\rm M_{\rm 3}$) and three magnitudes fainter. The integrated
 luminosity function, $\Phi({\rm M_{3}},{\rm M_{3}}+3,{\it z}_{\rm
   cl})$, is measured over the same luminosity range. We evolve the
 z=0.1 \citet{2003ApJ...592..819B} SDSS {\it r}-band luminosity
 function
 ($\phi_{\star}$=1.49$\times$10$^{-2}$,~M$_{\star}$=-20.44,~$\alpha$=-1.05)
 using the prescription described in \citet{1999ApJ...518..533L} that
 adds redshift-dependent terms to $\phi_{\star}$ and M$_{\star}$ with
 parameters P=-1.06 and Q=1.82. D, the angular diameter distance, is
   derived from the cluster redshift $z_{\rm cl}$. $\gamma$ and $\rm
   I_\gamma$ respectively define the slope of the angular galaxy
   correlation function and the integration constant arising from
   de-projecting the cluster. We set these to $\gamma=1.77$ and $\rm
   I_\gamma=3.78$. The correlation amplitude $A_{gc}$ is defined as:

\begin{equation}
\label{agc}
\rm A_{\rm gc}={\rm N_{\rm net}\over{\rm N_{\rm bg}}} {(3-\gamma)\over{2}} \theta^{\gamma-1}
\end{equation}

 where $\rm N_{\rm net}$ is the background-corrected count of galaxies
 within the luminosity range described above, out to an angular
 separation $\theta$ that corresponds to $0.5\mpc$ at the cluster
 redshift.  $\rm N_{\rm bg}$ is the background galaxy count within
 this radius, estimated from the mean density of galaxies across the
 whole field. The full 270\,deg$^2$ Stripe 82 catalogue provides
 additional definitions of cluster richness - we refer readers to
 GMB11 for the details of those measurements.

\subsection{Cluster sequence scatter ({\tt scatter})}
\label{scatter}
To estimate the width of a detected cluster's sequence, we first make
a fit to the slope of the sequence and remove the tilt. Using cluster
members between $m_{\rm BCG}\leq m \leq m_{\rm BCG}+3$, we estimate
the sequence scatter by making a $2\sigma$ clip in the colour
distribution.
 
The robustness of the red sequence fit is sensitive to the number of
members in the detection. Based on a bootstrap-resampling of the
cluster sequences, we find the fitting procedure is robust in clusters
with at least 8 members. Below this, sequence scatter estimates are
dominated by fitting uncertainty. For systems of at least 10 members,
the characteristic error in the sequence scatter is 34\%, dropping to
19\% for clusters with up to 30 members and 8\% for those with at
least 50 members. Future catalogues will provide improved estimates of
the sequence-fitting error.

\subsection{Projected scale ({\tt $\theta_{80}$}) \& concentration ({\tt $C$})}
\label{r80}
 For each cluster, a projected scale size $\theta_{80}$ is
 provided. This is calculated as the angular radius (in degrees)
 enclosing 80\% of cluster members from the centre.

 A measure of the projected concentration (C) is determined by
 comparing the radius enclosing 80\% of the cluster members to the
 radius enclosing 20\%. High values of $\theta_{80}/\theta_{20}$
 indicate a centrally concentrated cluster. 

\subsection{Testing the algorithm}
\label{sdss_tests}
\subsubsection{Cluster re-detection robustness}
\label{robustness}
 To determine how robust the detector is to catalogue incompleteness,
 we attempt re-detections of the Abell 2631 cluster after removing a
 random selection of members from the source data. Our sole constraint
 is that the cluster BCG remains in the source data. In the following
 analysis, we only consider the detected cluster closest to the
 original Abell 2631 position. Robustness is defined as the fraction
 of members detected in the new cluster from those remaining in the
 input catalogue. We use a test {\it g-r} photometric filter that
 adopts a $\beta_{g-r}$, $\cm$ and $\sigma_{g-r}$ best suited to the
 recovery of A2631, selecting approximately $85\%$ (108) of the
 visually selected members. We experiment with removal fractions down
 to $95\%$, corresponding to the largest fraction still retaining
 \nmin=5 original cluster members in the sample. 
\begin{figure}
\centerline{\includegraphics[width=0.5\textwidth]{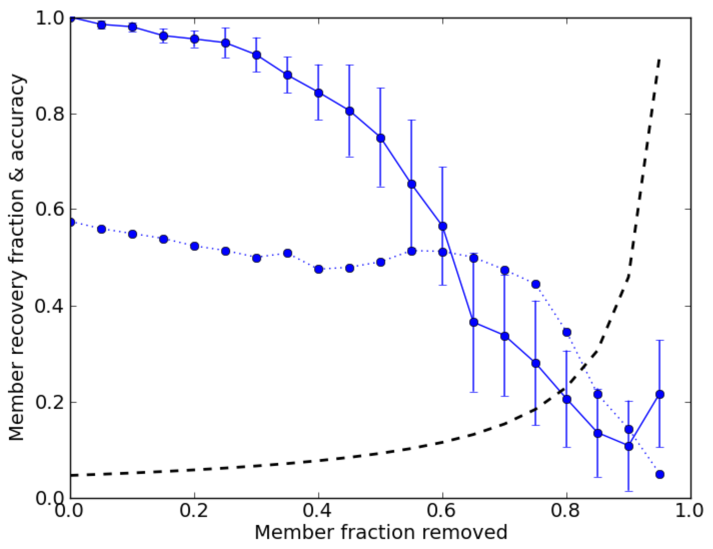}}
\caption{The recovery fraction (solid line) and recovery accuracy (dotted
  line). Some Abell 2631 cluster galaxies are randomly removed from
  the source catalogue, and the fraction subsequently identified in a
  re-detection of the cluster is the recovery fraction, with error
  bars of $1\sigma$ uncertainty calculated from 50 re-detections. The
  fraction of visually identified Abell 2631 galaxies making up the
  re-detected cluster defines the recovery accuracy. The fraction
  required to produce an \nmin=5 member system is denoted by the
  black dashed line.
\label{robust_fig}
}
\end{figure}

 Fifty random realisations of a depleted input catalogue are generated
 for each removal fraction, yielding a median recovery rate based on
 members that could have been added to the cluster. The solid blue
 line in Figure \ref{robust_fig} shows how increasing the removal
 fraction affects the fraction of cluster members recovered; error
 bars on this line represent $1\sigma$ uncertainties from the 50
 re-detections in each bin. The recovery fraction when no galaxies
 have been ejected is $\sim 93\%$ of the 108 A2631 members located
 inside the photometric filter. The other $7\%$ were rejected by the
 algorithm because either their Voronoi cells have insufficient
 densities to join the overdense collection of cells ($\probthresh$,
 see \S\ref{vt_parameters}), or their inclusion causes the
 percolating cluster to drop below the critical density (\sigmacrit).

 We take into account this intrinsic detection inefficiency, quoting
 yields from the cluster re-detection relative to the $\sim 93\%$ of
 members recovered where no additional galaxies are
 removed. Unsurprisingly, the fraction of detected members located in
 the cluster drops as more members are excised. However, over $75\%$
 of remaining members are re-detected even after half of the cluster
 is removed. Approaching larger removal fractions, the fragmentation
 of cluster members into spatially distinct groups hinders recovery of
 the complete set. The black dashed line in this plot corresponds to
 the minimum recovery fraction required to identify \nmin=5
 original members from the input data. The algorithm can robustly
 identify the original cluster down to an $80\%$ removal fraction,
 corresponding to 22 of the original 108 galaxies. Below this limit,
 an insufficient number of cluster members are recovered by the
 detector to identify a cluster associated with the halo.

 For each ejection fraction we also calculate the recovery accuracy:
 the fraction of visually identified A2631 galaxies making up the
 re-detected cluster.  The dotted blue line in Figure \ref{robust_fig}
 shows this parameter. The initial accuracy (no members are removed)
 is approximately 60\%, providing some estimate of our level of
 incompleteness when visually identifying cluster membership. As more
 members are removed, there is a gradual reduction in accuracy,
 implying replacement of these members with other galaxies becomes
 more commonplace. At large ($>70\%$) removal fractions, fragmentation
 acts to reduce the connectivity of cluster members, increasing the
 number of contaminant galaxies that share the photometric filter.

\begin{figure}
\centerline{\includegraphics[width=0.5\textwidth]{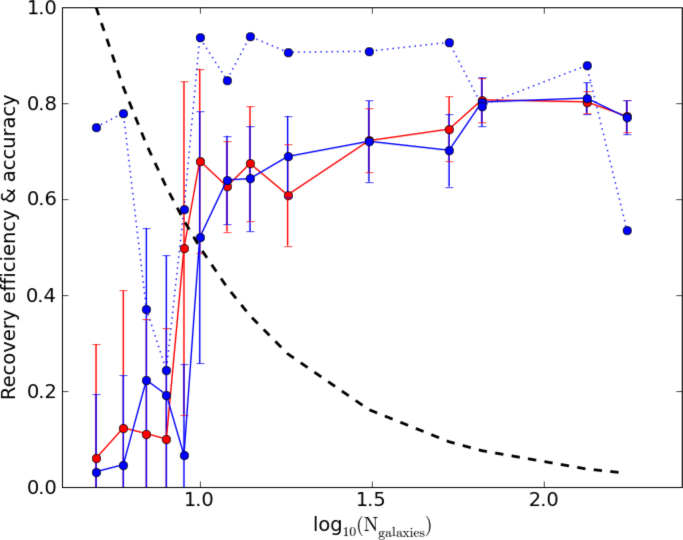}}
\caption{The algorithm's re-detection capability when a cluster
  has been moved to a random position. The recovery efficiency (solid
  blue line) is the fraction of original cluster galaxies found in the
  displaced cluster. The edge-effect recovery efficiency (red line)
  shows a similar test, instead moving the cluster to a random
  position near the survey boundary. Uncertainties in both lines are
  $1\sigma$ errors from 50 re-detections. The recovery accuracy
  (dotted blue line) is the ratio of input cluster members to the
  member count of the re-detected cluster. The black dashed line
  indicates the \nmin=5 threshold required to secure a robust
  detection of the cluster's halo.
\label{displacement_fig}
}
\end{figure}

\subsubsection{Cluster displacement and edge effects}
\label{displacement}
 A cluster detector should identify systems irrespective of the projected
 environment they are located in. Ideally then, recovery of identified
 members is achieved even if the system is moved to another position.

 To determine the sensitivity of cluster identification to localised
 background fluctuations, we shift source data positions of known
 cluster members to a random location, keeping their spatial
 distribution intact. A buffer is created around the survey edge to
 ensure no cluster members are displaced outside the boundaries, then
 a re-detection of the cluster is attempted. The re-detection
 performance is quantified by the recovery efficiency - the fraction
 of original members in the new cluster, and the recovery accuracy
 remains as defined in the previous test.

 Figure \ref{displacement_fig} shows the recovery efficiency ({\it
   solid blue}) and recovery accuracy ({\it dotted blue}) for clusters
   spanning more than an order of magnitude in membership (\nmin=5 to
   174 galaxies). If there was a choice of cluster for a membership
   bin, we used the system with the smallest sequence scatter to
   determine the impact of displacement on the best candidate in that
   membership group. Each cluster was re-detected in the pair of
   selection filters it was originally identified in, meaning a
   re-detection with no displacement would yield a perfect recovery
   efficiency and {\it recovery accuracy} (both equal to unity). We
   perform 50 random displacements for each of the selected clusters,
   using their scatter to derive 1$\sigma$ uncertainties from the
   mean. The black dashed line in Figure \ref{displacement_fig}
   corresponds to the recovery fraction required to detect \nmin=5
   galaxies of the original system from each displaced cluster.

 For the majority of cluster sizes, recovery accuracies are
 approximately constant at $\sim 90\%$, meaning $10\%$ of the cluster
 members are background galaxies selected in the same photometric
 selection. Recovery efficiency data suggest the detector makes
 significant cluster re-detections for systems down to 10 members, but
 smaller groups are susceptible to higher levels of contamination and
 fragmentation. Our example case of Abell 2631 (at
 $\rm log_{10}\rm N_{\rm gal}\sim 2.1$), with a recovery efficiency of 80\% is
 approximately 13\% lower than the recovery fraction from robustness
 test calculated above. A recovery accuracy of $\sim$86\% is
 consistent with the detector swapping 13\% of original members with
 background galaxies when the cluster is moved.

 We next establish how survey edges bias the detection of systems at
 the boundaries. Using the same set of clusters, we repeat the above
 experiment, specifically placing systems close to the survey edges to
 quantify the impact of edge effects on group and cluster
 recovery. When moving each cluster, we ensure no members are
 outside of the survey boundary. The average separation between survey
 edge and the member furthest from the cluster centre is around 23
 arcseconds.

 Galaxy cells at the boundary of a Voronoi Diagram are unbounded,
 often resulting in very large cell areas. This may hamper the
 identification of low-membership clusters, where a member with cell
 area exceeding the probability threshold may preclude the cluster
 from detection. Random positions are selected along any one of the
 four sides of the survey (allowing clusters to reside in a
 corner). In our source catalogue, the declination boundaries (at
 $\delta=\pm1.25^{\circ}$) are set by the geometry of the stripe,
 whilst the RA boundaries are artificially defined. Distances between
 the cluster centroid and survey edge are large enough to include all
 members within the survey. The red line in
 Figure \ref{displacement_fig} shows the recovery efficiency
 based again on 50 randomised displacements. This distribution is very
 similar to that of the displacement test above, suggesting edge
 effects do not hinder the recovery of clusters any more than the
 displacement of the members themselves. This is particularly
 significant at group scales, where the exclusion of one or two
 members could prevent the detection of the system. 

\subsubsection{False positive detection rate}
\label{colour_shuffle}
We set the detector the task of attempting to detect spatially
clustered systems with randomised colours. This establishes the
importance of red sequences to cluster detection with this algorithm
and provides an estimate of the false detection rate. We run the
detector on the source catalogue in the same manner as before, having
first shuffled the colours so while cluster members still reside in
high surface density regions, they no longer have red-sequences. We
identified two ``clusters'' (with 5 and 6 members) in the 7 square
degree survey, both located at the positions of original
high-membership \orca clusters. To ensure this calculation is
uninfluenced by the size of the survey, we repeat this process on the
full Stripe 82 dataset ($-50^{\circ}<\alpha<59^{\circ}$) covering
270 square degrees. The algorithm detects 15 ``clusters'' from these
data, each consisting of five or six-member groups. From this we infer
the number of spurious systems detected per 7 square degrees is 0.39.

In a similar fashion we next randomise galaxy positions while keeping
the colours the same. This means cluster red-sequences remain intact
as the algorithm scans through colour-magnitude space, but points
clustered in colour are no longer clustered in the sky. The algorithm
detected four ``clusters'' over the full 270 square degree Stripe 82
dataset, implying a $\sim$0.1\% spurious cluster detection rate.

Both exercises suggest the detector cannot identify clusters without
correlations in both colour and spatial position. Moreover the
probability of detecting systems based on random distributions of both
colour and position is below $1\%$.
 
\subsubsection{Projected cluster-pair resolution}
\label{projection}
 The ideal algorithm can identify two clusters with the same angular
 position on the sky, but at different radial distances. Using the
 $\cm-z$ relation demonstrated in Figure \ref{ri_z}, one
 can in principle isolate superimposed systems by identifying them in
 different filters. Within a detection filter $f(C_{\rm A})$ of width
 $\sigma_{f}$, two spatially coincident systems will be merged even if
 their sequences do not directly overlap. We overcome this limitation by
 splitting sequences in the following colour ($C_{\rm B}$) with the
 application of joint filters (\S\ref{sub_filter}). The resolving
 power of the algorithm in projection is therefore limited by the
 merging of separate clusters that are mistaken as multiple detections
 in \S\ref{merging}.

\begin{table*}
\begin{center}
\caption{An extract from the \citet{2007ApJ...660..239K} catalogue
  noting the 22 {\tt maxBCG} clusters within the limits of this SDSS
  sample field. The cluster name follows the IAU JHHMMSS+DDMM.m
  format. The RA and DEC are J2000, and measured in degrees. $z_{\rm
  photo}$ and $z_{\rm spec}$ are the estimated photometric and
  spectroscopic redshifts of the clusters. $\rm N_{\rm gal}$ is the number
  of members in the cluster, and $\rm N^{\rm R200}_{\rm gal}$ is the
  scaled richness.}
\begin{tabular}{|c|c|c|c|c|c|c|}
\hline
Cluster name & RA & DEC & $z_{\rm photo}$ & $z_{\rm spec}$ &
$\rm N_{\rm gal}$ & $\rm N^{\rm R200}_{\rm gal}$ \\
\hline
BCG J233740+00160.3 & $354.41553$ & $0.27138$ & $0.286$ & $0.277$ & $59$ & $88$ \\
BCG J234624+00440.0 & $356.59955$ & $0.74943$ & $0.273$ & $0.275$ & $25$ & $26$ \\
BCG J233746-00420.2 & $354.44067$ & $-0.70310$ & $0.286$ & $0.287$ & $20$ & $17$ \\
BCG J234100+00040.9 & $355.24905$ & $0.08161$ & $0.194$ & $0.185$ & $23$ & $23$ \\
BCG J233955-00250.0 & $354.97916$ & $-0.43282$ & $0.275$ & $0.277$ & $17$ & $15$ \\
BCG J234548-01070.7 & $356.45068$ & $-1.12775$ & $0.273$ & $-$ & $18$ & $18$ \\
BCG J234604-00100.0 & $356.51477$ & $-0.18283$ & $0.254$ & $-$ & $22$ & $22$ \\
BCG J234322+00190.6 & $355.84039$ & $0.32587$ & $0.257$ & $0.267$ & $38$ & $60$ \\
BCG J234146+01070.5 & $355.44077$ & $1.12444$ & $0.246$ & $0.251$ & $15$ & $11$ \\
BCG J233919-00150.6 & $354.82941$ & $-0.25941$ & $0.284$ & $-$ & $14$ & $11$ \\
BCG J234024-00050.6 & $355.10205$ & $-0.09300$ & $0.281$ & $-$ & $17$ & $13$ \\
BCG J234720+00290.7 & $356.83487$ & $0.49456$ & $0.286$ & $0.275$ & $12$ & $10$ \\
BCG J233900+00420.0 & $354.75143$ & $0.71610$ & $0.219$ & $0.183$ & $14$ & $11$ \\
BCG J234122+00190.0 & $355.34253$ & $0.33330$ & $0.284$ & $0.278$ & $22$ & $22$ \\
BCG J233911-01130.3 & $354.79459$ & $-1.22236$ & $0.292$ & $-$ & $14$ & $10$ \\
BCG J234626+00430.7 & $356.60690$ & $0.72794$ & $0.251$ & $-$ & $25$ & $29$ \\
BCG J234403+00130.6 & $356.01273$ & $0.22646$ & $0.262$ & $-$ & $16$ & $11$ \\
BCG J234233-00170.3 & $355.63776$ & $-0.28873$ & $0.275$ & $-$ & $16$ & $14$ \\
BCG J233755+00130.5 & $354.47760$ & $0.22478$ & $0.262$ & $0.278$ & $37$ & $61$ \\
BCG J233825-00090.2 & $354.60291$ & $-0.15397$ & $0.270$ & $-$ & $14$ & $11$ \\
BCG J234737-00370.9 & $356.90375$ & $-0.63221$ & $0.262$ & $-$ & $14$ & $11$ \\
BCG J234106+00120.4 & $355.27640$ & $0.20707$ & $0.262$ & $-$ & $15$ & $10$ \\
\hline
\end{tabular}
\label{maxBCG_table}
\end{center}
\end{table*}

We test this effect with the same clusters used in
\S\ref{displacement} by implanting a 7-member test cluster at the
same spatial position and colour normalisation $\cm$.  We increase the
test cluster $C_{\rm B}$ colour normalisation by $\delta\cm$ and run
the matching algorithm. This is repeated until the detector classifies
the reddened test cluster as an independent system. The {\it resolving
  capability} of the algorithm can be parametrised as $\chi=
\Delta\cm/\sigma$: the minimum sequence colour separation between the
two detected systems relative to the width of the filters they were
identified in. Small values indicate a good resolution, and in all
clusters tested against, we found $\chi<0.5$. Moreover, for all but
two membership bins ($\rm N_{\rm gal}$=14,18) the test cluster was resolved
within $\chi<0.25$. Whilst in our real astronomical data we observe
some cluster pairs overlapping in projected space, these examples
exhibit large separations in both colour space and redshift. For
example the two clusters {\it MGB J234729-00080.4} and {\it MGB
  J234733-00100.0} have redshifts of $z=0.23$ and $z=0.53$ and
$\chi_{r-i}=7.8$. Although our analysis here could benefit from a
larger sample size, \orca can distinguish between two separate systems
even if their sequences lie in the same filter, subject to their
colour separation being at least 1/4 the filter width. Below this
level, their similarity in colour likely justifies classifying these
systems as the same structure.

\section{Comparison to existing cluster data}
\label{comparisons}
 The positions of detected clusters can be seen in Figure
 \ref{s82_field}, with the location of {\tt maxBCG} clusters
 \citep[][]{2007ApJ...660..239K} marked with red circles, and the
 positions of known X-ray clusters marked with blue squares. Clusters
 detected in the \{{\it g-r}, {\it r-i}\} combination are shown as
 blue filled cells, those detected in \{{\it r-i}, {\it i-z}\} are red
 filled cells. In each case the cluster BCG cells are yellow.

\subsection{The maxBCG catalogue}
\label{maxBCG}
\begin{figure*}
\centerline{\includegraphics[width=1.0\textwidth]{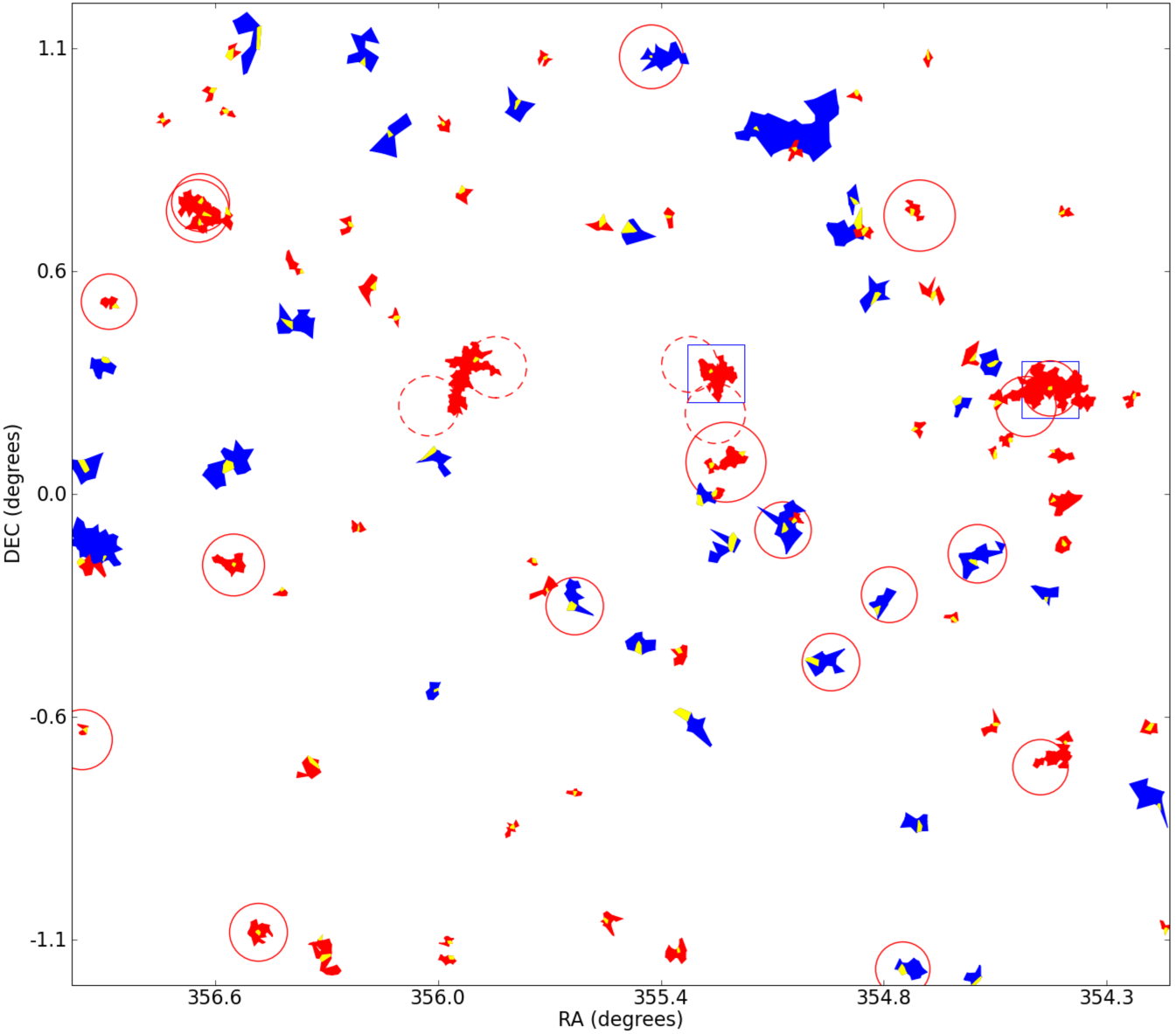}}
\caption{Clusters detected in the Stripe 82 field. The coloured cells
  represent clusters detected in different colour pairs. Blue cells
  correspond to clusters detected in \{{\it g-r}, {\it r-i}\} filter
  pairs, red clusters detected in \{{\it r-i}, {\it i-z}\} filter
  pairs. Yellow cells indicate the BCG position of each cluster. Red
  circles indicate the position of {\tt maxBCG} clusters, based on data
  shallower than that used in the study here. Circle radii correspond
  to $1h^{-1}Mpc$, based on the {\tt maxBCG} photometric redshift estimate
  of the cluster. Dashed red circles indicate the four {\tt maxBCG}
  clusters discussed in \S\ref{maxBCG} that also feature {\it
    gri}-colour imaging in Figures \ref{filament} and
  \ref{MGB_J234105+00180.3}. Blue squares note the position of
  ROSAT All Sky Survey X-ray sources, with half-lengths corresponding
  to $1h^{-1}Mpc$.
\label{s82_field}
  }
\end{figure*}

 The \citet{2007ApJ...660..239K} {\tt maxBCG} catalogue of 13,823
 optically selected SDSS clusters uses the detection algorithm
 described in \citet{2007ApJ...660..221K}. This catalogue makes use of
 data from an earlier release of SDSS, so was unable to take advantage
 of the added depth Stripe 82 offered this study. Because direct
 comparison of the two cluster selection functions is both non-trivial
 and unfair, we do not attempt a full analysis in this study. However,
 in the spirit of matching detections made here to those of the
 shallower data in the \citet{2007ApJ...660..239K} catalogue, we
 include the positions of {\tt maxBCG} clusters in Figure
 \ref{s82_field} as a set of red circles. The centre of these circles
 is the location of the assigned Brightest Cluster Galaxy (BCG),
 whilst the radius corresponds to 1$\mpc$ calculated from the
 published photometric redshift estimate of the cluster. We stress
 however, that this does not necessarily correspond to the physical
 size of the cluster.

 The survey area contains 22 {\tt maxBCG} clusters. For ease of
 reference, salient details from that catalogue are reproduced in
 Table \ref{maxBCG_table}, along with a name of the form BCG
 JHHMMSS+DDMM.m. We attempt a simple match to the \orca catalogue by
 looking for either common BCGs (and more generally a match to \orca
 cluster members where BCGs are assigned differently) or statistically
 significant separations between \orca centroids and {\tt maxBCG}
 positions. We find a match to 18 of the 22 clusters; the four {\tt
   maxBCG} clusters that do not have \orca analogues are noted in
 Figure \ref{s82_field} with dashed circles and are apparent as two
 pairs with small angular separation.

 We note the \orca cluster ({\it MGB J234341+00180.3}) is situated
 between the western pair ({\it BCG J234322+00190.6} and {\it BCG
   J234403+00130.6}). Optical-band imaging (Figure \ref{filament} in
 Appendix) shows evidence of early type galaxies distributed in a
 filamentary chain, approximate comoving length $2\mpc$, sampled by
 \orca between the {\tt maxBCG} detections.

\begin{figure*}
\centerline{\includegraphics[width=1.0\textwidth]{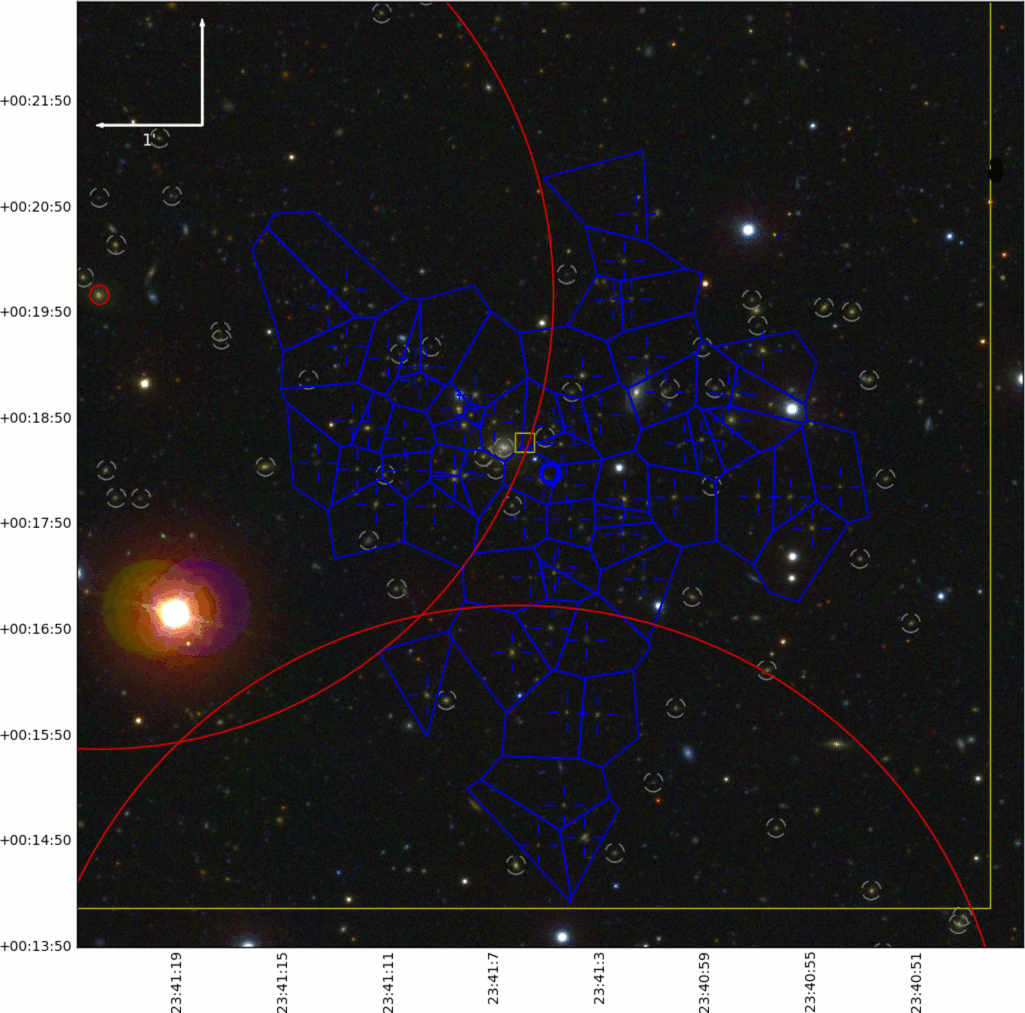}}
\caption{Stripe 82 cluster {\it MGB J234105+00180.3}: an \orca
   detection between two {\tt maxBCG} clusters and on top of an X-ray
   cluster position. Members and their Voronoi cells are marked in
   blue, the thick circle indicating the estimated cluster
   centre. Grey dashed circles are {\it associate cluster members}
   arising from multiple detections of this cluster
   (\S\ref{merging}). Red data indicate the location of {\tt maxBCG}
   clusters {\it BCG J234122+00190.0} and {\it BCG J234106+00120.4},
   with larger circles indicating a 1$\mpc$ radius, smaller circles
   the BCG positions. Yellow data indicate the NORAS X-ray cluster
   {\it RXC J2341.1+0018}; the half-length of the large square
   corresponds to 1$\mpc$ based on the cluster redshift, the small
   square noting the X-ray position, uncertain to approximately
   1'. The X-ray-\orca centroid separation is approximately $0.4'$.
\label{MGB_J234105+00180.3}
  }
\label{lastpage}
\end{figure*}

 The other pair ({\it BCG J234106+00120.4} and {\it BCG
   J234122+00190.0}) may be part of an elongated structure sampled by
 both the four {\tt maxBCG} entries in that area and also by the \orca
 detector. Figure \ref{MGB_J234105+00180.3} shows the \orca cluster
 MGB J234105+00180.3. This cluster centre, situated between the two
 {\tt maxBCG} clusters, matches the centroid of an RASS cluster to
 within $0.4'$, with an uncertainty of $\sim 1'$ in the X-ray source.

 Overall, we find very good agreement with the {\tt maxBCG} catalogue
 of clusters, detecting 81\% of their entries in the survey region,
 rising to 100\% when taking into account how the different algorithms
 handle systems that by eye resemble filamentary structure.

\subsection{X-ray detected clusters} 
\label{rass} 
X-ray selected cluster catalogues are useful independent checks on the
population of clusters detected by optical cluster-finders. We use
cluster data from the {\it ROSAT} All Sky Survey-derived
\citep[RASS;][]{1999A&A...349..389V} NORAS \citep{2000ApJS..129..435B}
and BCS catalogues \citep[for the latter, both main and extended
  catalogues;][]{1998MNRAS.301..881E,2000MNRAS.318..333E}, the XCS
\citep{2001ApJ...547..594R,2011arXiv1106.3056M} and BLOX
\citep{2007A&A...470..821D} from {\it XMM-Newton}, and CHaMP
\citep{2006ApJ...645..955B} from {\it Chandra}. We combine these
datasets, taking care to identify any duplicate detections, to form an
X-ray catalogue consisting of 1463 unique clusters. From this
catalogue there are 58 X-ray clusters within the full 270 square
degree footprint covered by Stripe 82, and two of these lie within the
7 square-degree sample studied here. In future we will provide a
comparison of these X-ray data to an optical cluster catalogue
covering a larger area.

Blue squares in Figure \ref{s82_field} show the position of the two
clusters in the region we study here. The westernmost X-ray cluster,
{\it RXC J2337.6+0016} \citep[also detected in the flux-limited
Brightest Cluster Sample,][]{1998MNRAS.301..881E} is the X-ray
counterpart to {\it ACO2631} \citep{1989ApJS...70....1A} and has a
redshift of 0.2780 \citep{1995MNRAS.274...75C}. The X-ray position
coincides with the \orca detection of this system ({\it MGB
J233740+00160.2}; z=0.2571) at a separation ($\Delta\theta,\Delta \rm
z$) of ($0.1',0.021$). The easternmost X-ray cluster ({\it RXC
J2341.1+0018}) with a redshift of z=0.2766 \citep[][misidentified as
{\it ACO2644}]{1998A&AS..129..399K} was originally optically
identified in \citet{2002AJ....123.1807G} and
\citet{2004AJ....128.1017L}, and is in close proximity to {\it MGB
J234105+00180.3} (z=0.2588), with ($\Delta\theta,\Delta \rm
z$)=($0.4',0.018$). This latter match also appears to straddle two
{\tt maxBCG} clusters in the same region as the potentially elongated
structure discussed in \S\ref{maxBCG}.

\section{PS1 Mock cluster catalogue}
\label{mocks}
\subsection{Simulations}
\label{mock_intro}
\begin{figure*}
\centerline{\includegraphics[width=1.00\textwidth]{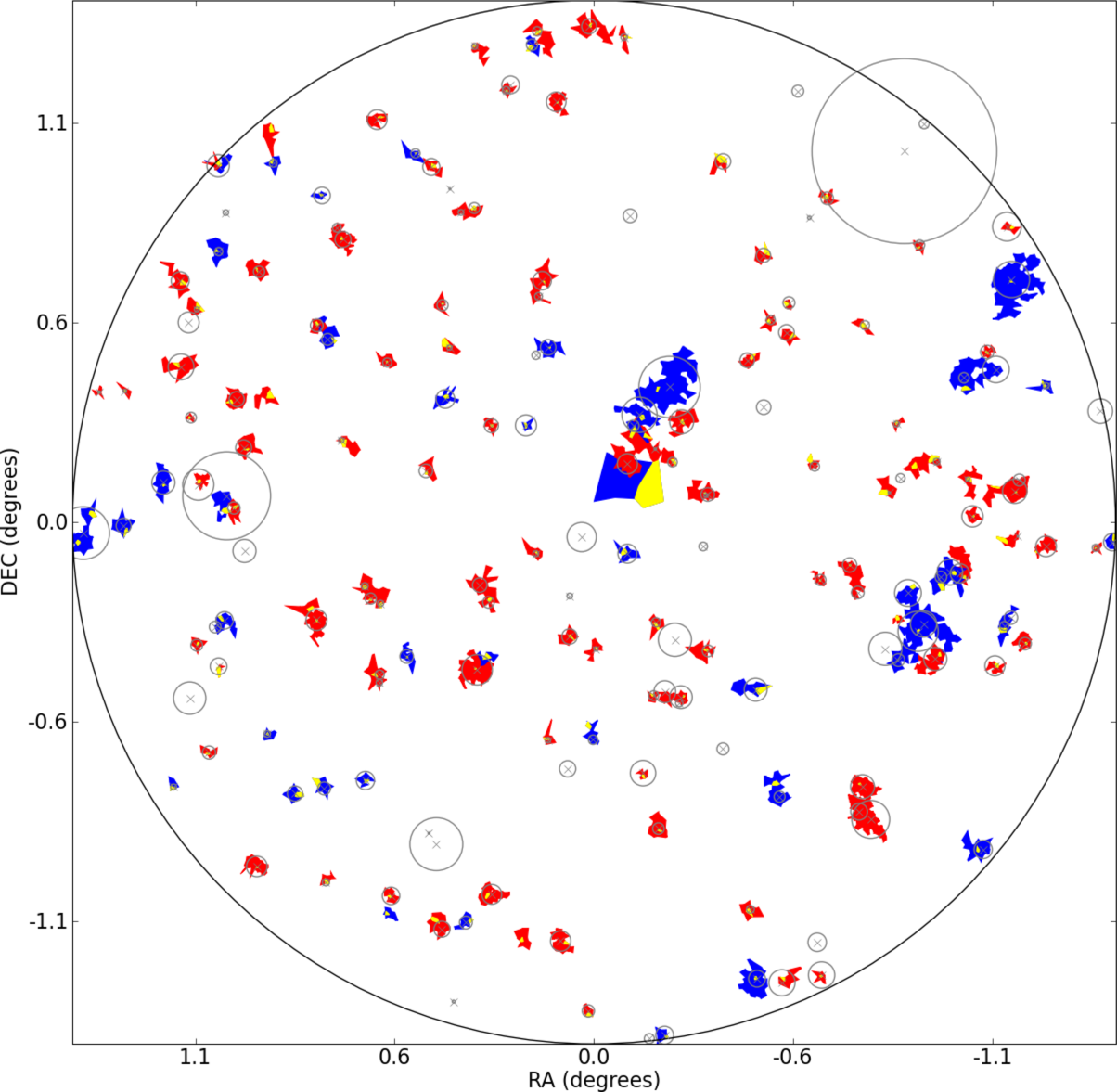}}
\caption{Clusters in haloes of mass $\geq 10^{13.5}\mass$ from the mock
  \orca cluster catalogue (cells) and the \lcdm catalogue
  (circles). Cell colours correspond to clusters detected in different
  colour pairs. Blue cells are clusters detected in the \{{\it g-r},
  {\it r-i}\} filter pairs, red are clusters detected in \{{\it r-i},
  {\it i-z}\}. Yellow cells indicate the BCG of each cluster. Crosses
  denote the \lcdm cluster centre, and circle radii indicating the
  angular distance between the centre and most distant member.
\label{mock_field}
}
\end{figure*}
In this section, we describe the application of \orca to a mock PS-1
lightcone. Theoretical simulations allow one the luxury of comparing
clusters detected by the algorithm (\orca clusters) to the galaxy
membership of dark matter haloes (hereafter \lcdm clusters). Simulated
galaxies are allocated to dark matter haloes using the
\citet{2006MNRAS.370..645B} semi analytic model. This approach makes
the assumption a satellite galaxy is stripped of hot gas immediately
following accretion onto a large halo. Star formation is halted after
the cold gas reservoir is depleted, and the galaxy joins the red
sequence. Coupled with AGN feedback, this prescription reproduces the
observed bimodality in galaxy colours. However a known flaw, the rate
of gas depletion, results in redder than observed satellite
galaxies. Recent treatments of ram-pressure stripping
\citep[e.g.,][]{2008MNRAS.383..593M} hope to improve understanding of
the transition to early-type galaxies with improved semi-analytic
models \citep{2008MNRAS.389.1619F,2010MNRAS.405.1573B}.

Although mock surveys are inaccurate realisations of the
universe \citep[see][for an example in a cluster detection
context]{2010MNRAS.404..486H}, they can nevertheless serve as
self-consistent tests of the detector. We emphasise, however, there is
little merit in comparing mock cluster detections with those in survey
data until models can reproduce the observed group and cluster galaxy
population with more fidelity.

To compare \orca detections to the model, we construct \lcdm clusters
with the aid of halo memberships and full 3D galaxy data. In each
\lcdm cluster, we calculate the approximate centre from cluster member
positions. Outlier galaxies are identified by rejecting 3$\sigma$
deviations from a bootstrap-estimated median galaxy-centroid
distance. Following outlier ejection, we find the resultant cluster
sizes agree well with the virial radii of the host haloes. We set a
minimum cluster mass limit by selecting \lcdm clusters residing in
haloes with $\rm M_{\rm H}\geq 10^{13}\mass$.

\subsection{Mock reference cluster}
\label{mock_trainer}
We select a ``reference cluster'' from a set of $\Lambda$CDM-based
detections generated from a preliminary scan of the simulation. The
chosen cluster allows us to set the slope and width of the photometric
filters in our search through the mock data. Candidate training
clusters were identified from a redshift range bracketing Abell 2631
($z=0.278$), with similar memberships and a clear sequence in all
colours. We selected the richest of these candidates, featuring 130
members and a redshift of $z=0.3$. By applying the same fitting
techniques as those described in \S\ref{filter_parameters}, we set the
filter parameters listed in Table \ref{mock_filter_table} and apply
the same colour ranges as those used on the SDSS. The fitted gradients
are steeper in {\it g-r} and {\it r-i} than those used for the SDSS,
and the filter widths are smaller. These values were nevertheless
consistent with the other candidate reference clusters identified in
the mock. As before, we use the most conservative width ({\it g-r},
0.13) for filters in each colour.

\begin{table}
\begin{center}
\caption{Filter parameters fitted from the mock reference cluster (by
  analogy with those derived from Abell 2631) along with colour ranges
  searched by the detector (the same as those used in the Stripe 82
  data).}
\begin{tabular}{|c|c|c|c|c} \hline
Colour & Slope ($\beta$) & Width ($\sigma$) & Range & Filters\\
\hline
{\it g-r} & $-0.070$ & ${\bf 0.130}$ & $0.47-2.00$ & $39$ \\
{\it r-i} & $-0.032$ & $0.064$ & $0.00-1.22$ & $38$ \\
{\it i-z} & $-0.012$ & $0.035$ & $-0.10-1.10$ & $31$ \\
\hline
\end{tabular}
\label{mock_filter_table}
\\
\end{center}
\end{table}

\subsection{Producing \lcdm and mock \orca cluster catalogues}
\label{mock_catalogue}
 Except for the revised parameters listed in Table
 \ref{mock_filter_table}, the detector ran as described in
 \S\ref{method}, and applied magnitude limits created a source
 catalogue of 80,536 mock galaxies. Because the algorithm relies on
 the detection of colour-magnitude ridgelines, we do not want to
 include \lcdm clusters without detectable sequences. We therefore
 construct the \lcdm cluster list from galaxies selected in the same
 photometric filters used by the detector, meaning \lcdm clusters may
 also be detected multiple times. We group together \lcdm clusters
 with common halo identifiers, but as before selected the highest {\it
 reduced flux} candidate as the ``best'' \lcdm cluster.

We found a total of 305 \orca clusters with
$\rm M_{\rm H}\geq 10^{13}\mass$; at $\rm M_{\rm H}\geq 10^{14}\mass$ the counts are
more equal. Although the majority of clusters identified are at
$z\sim0.3$, the tests we describe in the following section will
highlight how well \orca performs over this entire parameter
space. Figure \ref{mock_field} shows a simple comparison of the
two catalogues by plotting both sets of clusters residing in haloes
$\rm M_{\rm H}\geq 10^{13.5}\mass$ out to $z=0.6$ (the highest cluster
redshift in the SDSS cluster catalogue). Grey circle centres denote
the position, and their radii the maximum member-cluster centre
distance of \lcdm clusters.  Blue and red cells represent \orca
clusters detected in \{{\it g-r}, {\it r-i}\} and
\{{\it r-i}, {\it i-z}\} respectively.

\subsection{Performance of the algorithm}
 To determine how well the detector recovers and characterises the
 mock clusters, we illustrate here three simple tests to quantify the
 detection performance.

\subsubsection{Completeness}
\label{completeness}
We define completeness as the number of detected haloes as a function
 of halo mass and redshift. A halo is detected if at least
 \nmin~galaxies are identified, even if they are shared between
 multiple \orca clusters (for example, fragmenting a halo when the
 algorithm attempts to identify substructure). We compare this number
 to \lcdm cluster counts (by definition unfragmented), with at least
 \nmin~members.

\begin{figure}
\centerline{\includegraphics[width=0.5\textwidth]{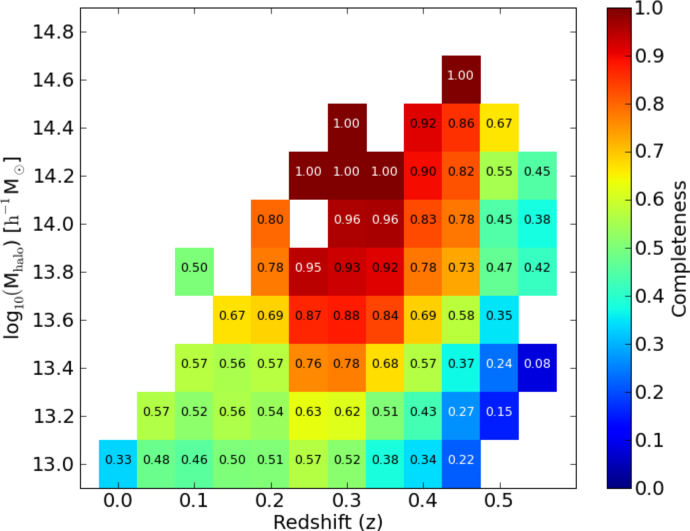}}
\caption{Completeness of mock \lcdm clusters. The fraction of
correctly detected clusters from the \orca catalogue as a function
of halo mass and redshift. The white regions indicate where there were
no \lcdm clusters in that bin.
\label{completeness_figure}
  }
\end{figure}

The fraction of detected \lcdm clusters can be seen in Figure
\ref{completeness_figure}, where we produce a grid of cells with
sampling intervals of 0.05 in redshift and 0.2 in ${\rm log_{10}}$
halo mass. Because in some cases only a few detections occupy each
cell, some regions will suffer from shot noise. We smooth the data
using a $3\times3$ grid so the completeness for a given cell is the
mean completeness over this region. Empty regions in Figure
\ref{completeness_figure} therefore indicate where either no
\lcdm clusters exist or too few clusters are found to reliably
calculate the completeness (we set a threshold of at least five
clusters detected over the $3\times3$ grid). Between $0.1\leq
z\leq0.4$, the detector attains at least 68\% completeness for halo
masses above $10^{13.6}\mass$, and is over 90\% complete in halo
masses exceeding $10^{14.3}\mass$. This compares favourably with the
{\tt maxBCG} algorithm applied to mock simulations, where
\citet{2007ApJ...660..221K} report $>90\%$ completeness between
$0.1\leq z\leq0.3$ for $\rm M_{\rm H}\geq 10^{14.3}\mass$ with clusters
containing at least 10 members (cf. \nmin=5 in this
study). Applying the completeness definition and the same selection
criteria as that study, the \orca detector is $>90\%$ complete down to
a halo mass of $10^{13.8}\mass$. These results also compare well to
the Voronoi Tessellation completeness of the {\tt 2TecX}
\citep{2009MNRAS.395.1845V} algorithm, either matching or exceeding
their stated completeness for $\rm M_{\rm H}=10^{13.7}$ and $10^{14}\mass$ up
to our redshift limit.

At higher redshifts there is a decline in completeness where there are
only a few members brighter than the magnitude limit, reducing the
algorithm sensitivity to distant clusters. This effect is more
apparent among the lower mass haloes. At high redshift ($z>0.4$) and
low mass ($\rm M_{\rm H}\leq10^{13.3}\mass$) there are 12 \lcdm clusters, but
the detector identifies only two of these. We also note a local
incompleteness at $z\leq0.08$. Arising from our choice of probability
threshold ($\probthresh$), too few overdense cells are selected in
filters featuring low signal-to-noise clusters. The photometric
filters best suited to detecting local, relatively blue clusters have
galaxy populations dominated by the blue cloud component of the
colour-magnitude relation. Successful detections in this crowded field
are compounded by the larger scale-size of more local clusters such as
the local ($z=0.03$) seven-member group at the north-western boundary
of the catalogue in Figure \ref{mock_field}. Under these
circumstances, it becomes unlikely cluster Voronoi cells share common
vertices, restricting potential membership links between them.

We classify spurious detections in the mock cluster catalogue as those
clusters where each member belongs to a different halo. Of the 305
\orca clusters, only two fit this description, suggesting a spurious
detection rate (0.7\%) consistent with tests performed in
\S\ref{colour_shuffle}.

\subsubsection{Stellar mass accuracy}
\begin{figure}
\centerline{\includegraphics[width=0.5\textwidth]{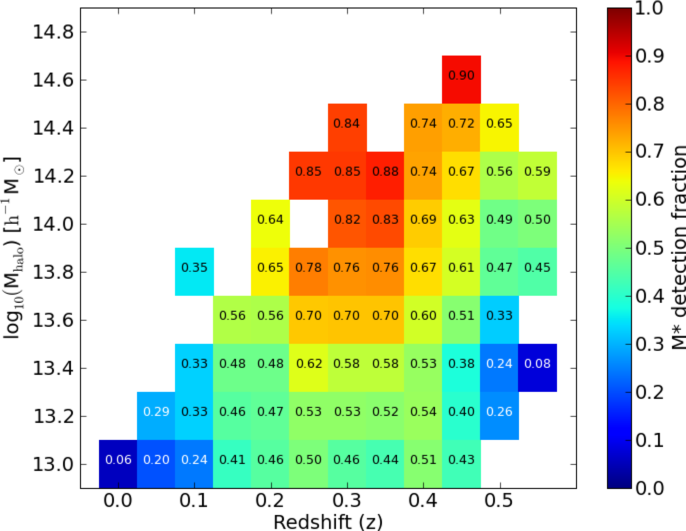}}
\caption{Stellar mass accuracy. The fraction of recovered stellar mass 
in mock clusters as a function of halo mass and redshift.
\label{sm_acc}
  }
\end{figure}

Stellar mass accuracy is the stellar mass of an \orca cluster relative
to that of the \lcdm cluster belonging to the same halo. Because the
algorithm may split the halo galaxies into multiple clusters, we
combine the mass of all \orca clusters sharing the same halo. In \lcdm
clusters with up to $\sim$12 members (approximately 75\% of the
catalogue), over half of the total cluster stellar mass comes from the
two most massive galaxies. The efficient detection of these galaxies
is therefore essential in gaining accurate estimates of cluster
stellar masses. The stellar mass accuracy for each \lcdm cluster is
$\rm A_{*}=\rm M_{*}^{\rm cl}/\rm M_{*}^{\rm true}$, where
$\rm M_{*}^{\rm cl}$ is the stellar mass of all \orca cluster
members registered to the \lcdm cluster's halo. We apply the same
gridding technique discussed in the previous section, requiring at
least 5 clusters in a grid to define a reliable $\rm A_{*}$. As Figure
\ref{sm_acc} shows, between $0.1\leq z\leq0.4$ the algorithm recovers
over half of the cluster stellar mass for systems with halo masses of
at least $10^{13.4}\mass$. This recovery fraction improves with
increasing mass, reaching 90\% in some cases. Both local and distant
clusters suffer from lower stellar mass estimates. For the former,
higher levels of halo fragmentation (one halo being assigned to many
\orca clusters) result in galaxies lost to nearby systems with
densities or memberships too low to qualify as clusters. Those systems
with redshifts $z>0.5$ tend to be unfragmented but contain fewer
members, causing an underestimation of cluster stellar mass. The
stellar mass accuracy at the median redshift of the survey ($z=0.33$)
remains above 50\% down to halo masses of $10^{13.2}\mass$, and above
75\% from masses of $10^{13.8}\mass$, suggesting the detector performs
well in estimating the true cluster stellar mass content.

\subsubsection{Purity}
\begin{figure}
\centerline{\includegraphics[width=0.5\textwidth]{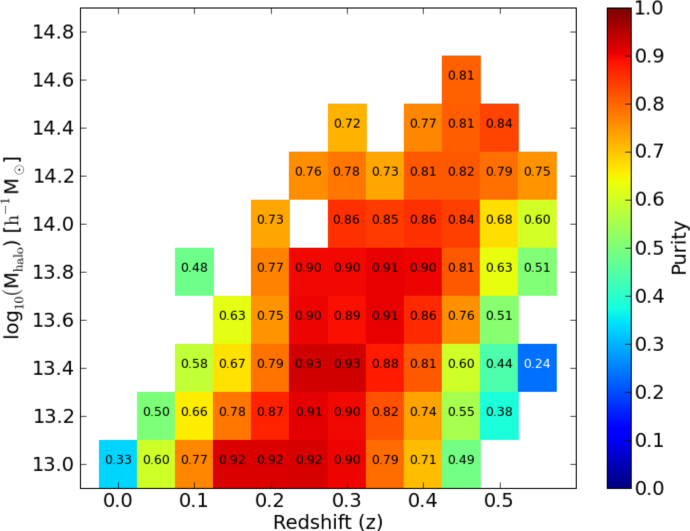}}
\caption{The purity of \lcdm clusters detected by the \orca
  algorithm. Low values indicate where clusters have included a large
  number of contaminating galaxies not belonging to the halo.
\label{purity_plot}
  }
\end{figure}

As discussed in \S\ref{completeness}, a halo is detected by the
algorithm if it finds at least \nmin~members that have been
allocated to \orca clusters. For a cluster with 7 members, the
distinction between a cluster containing 5 halo galaxies and 2
interlopers and one containing 7 halo galaxies provides a measure of
cluster purity. We define purity as the fraction of galaxies \orca
assigned to the cluster that are members additionally belonging to the
host halo. This description is in line with the purity described by
\citet{2007ApJ...660..221K}. However, we decide not to adopt a
threshold above which a cluster is considered pure, instead directly
assigning each cluster a purity fraction. Figure
\ref{purity_plot} shows the purity of \orca clusters with varying
redshift and halo mass, the gridding method here being the same scheme
introduced in \S\ref{completeness}. \orca clusters are at least
$70\%$ pure at the median redshift of the survey over all halo
masses. The purity appears to drop at higher redshifts, attributed to
faint but genuine cluster members being replaced by brighter
contaminants that lie on the cluster sequence. Relative to the
completeness and stellar mass estimates, cluster purity is not as
sensitive to halo mass. This is most likely a consequence of the
membership incompleteness discussed in
\S\ref{vt_parameters}. Because peripheral members are less likely
to be in Voronoi cells tagged as statistically significant, the
inclusion of interlopers at cluster edges is reduced. As in the
previous section, increased halo fragmentation drives the local drop
in purity, serving to increase the contamination fraction by
distributing the halo galaxies among local clusters and systems
failing to achieve cluster status.

\section{Summary}
\label{summary}
 We present and demonstrate a new cluster detection algorithm based on
 red-sequence cluster searches, the detection of overdensities using
 Voronoi Tessellations, and connecting galaxies into clusters with a
 Friends-of-Friends algorithm. With this approach, we make only two
 assumptions about the systems we are looking for: that they have
 detectable red-sequences, and are overdensities in the projected
 plane of the sky.

 We calibrate the photometric selection filters to a rich Abell
 cluster found in SDSS data, and find that recovery of members from
 both this large cluster and a small group is largely insensitive to
 the choice of two algorithm parameters controlling the behaviour of
 the algorithm. When applying the algorithm to a sample of SDSS Stripe
 82 galaxies with four bands, we find 97 clusters. Based on
 spectroscopic and photometric redshifts, we estimate these clusters
 are detected out to $z=0.6$ and the catalogue has a median redshift
 of $z=0.31$. We perform false-positive tests suggesting the spurious
 detection frequency is below $1\%$. Tests on the catalogue suggest
 the detector is robust to sparsely sampled cluster fields and is not
 overly sensitive to survey edges. In comparing our data to existing
 optical and X-ray clusters, we find good agreement with the {\tt
   maxBCG} and RASS catalogues in the same region.

 We go on to test the performance of the detector with a mock survey
 generated from a semi-analytic galaxy formation model. In comparing
 the \orca cluster detections to those generated from halo membership
 data, we make a quantitative assessment of the detector
 performance. The algorithm identifies 305 clusters, whilst the
 simulation produces 414 down to a halo mass of $10^{13}\mass$. At the
 median redshift of the catalogues (both $z=0.33$) we find \orca is
 75\% complete down to a cluster halo mass of $10^{13.4}\mass$ and is
 able to recover approximately 75\% of the total stellar mass for
 clusters in haloes of at least $10^{13.8}\mass$. 

 We have demonstrated this algorithm is capable of identifying
 clusters in both real and simulated data with minimal assumptions as
 to the nature of clusters. In combining comprehensive colour scans to
 search for cluster red-sequences with Voronoi diagrams to estimate
 surface densities, we avoid making model-dependent decisions about
 what a cluster is. Cluster redshifts arise as a consequence, not
 condition, of our detection, affording additional freedom from model
 SEDs and the uncertainties inherent in photometric redshift data
 spanning the depths, fluxes and areas set to be commonplace in
 next-generation galaxy catalogues. This detector can be used in any
 survey where there are at least two photometric bands, but is most
 powerful when applied to multi-colour surveys such as the forthcoming
 Pan-STARRS surveys. The scope for cluster detection with \orca is not
 limited solely to the optical regime. Preliminary tests with
 optical-IR band-merged catalogues show great promise, requiring
 minimal adaptation to facilitate the detection of the 4000\AA\ break
 into the IR bands and beyond $z=1$.

\section*{Acknowledgements}
We thank the referee for their useful comments which improved the
clarity of the paper. DNAM acknowledges an STFC PhD studentship, JEG
and RGB thank the U.K. Science and Technology Facilities Council and
the Natural Sciences and Engineering Research Council of Canada for
financial support. The authors thank Alastair Edge, John Lucey, Kathy
Romer, Ian Smail and John Stott for useful discussions, Carlton Baugh,
Yan-Chuan Cai and Shaun Cole for access to the mock MDS lightcone
data.

Calculations in portions of this work were performed on the ICC
Cosmology Machine, which is part of the DiRAC Facility jointly funded
by STFC, the Large Facilities Capital Fund of BIS, and Durham
University.

Funding for the SDSS and SDSS-II has been provided by the Alfred
P. Sloan Foundation, the Participating Institutions, the National
Science Foundation, the U.S. Department of Energy, the National
Aeronautics and Space Administration, the Japanese Monbukagakusho, the
Max Planck Society, and the Higher Education Funding Council for
England. The SDSS Web Site is http://www.sdss.org/.

The SDSS is managed by the Astrophysical Research Consortium for the
Participating Institutions. The Participating Institutions are the
American Museum of Natural History, Astrophysical Institute Potsdam,
University of Basel, University of Cambridge, Case Western Reserve
University, University of Chicago, Drexel University, Fermilab, the
Institute for Advanced Study, the Japan Participation Group, Johns
Hopkins University, the Joint Institute for Nuclear Astrophysics, the
Kavli Institute for Particle Astrophysics and Cosmology, the Korean
Scientist Group, the Chinese Academy of Sciences (LAMOST), Los Alamos
National Laboratory, the Max-Planck-Institute for Astronomy (MPIA),
the Max-Planck-Institute for Astrophysics (MPA), New Mexico State
University, Ohio State University, University of Pittsburgh,
University of Portsmouth, Princeton University, the United States
Naval Observatory, and the University of Washington.

\bibliography{./bibliography.bib}

\onecolumn
\newpage
\clearpage

\noindent
\begin{figure}
\begin{minipage}{\linewidth}
\section*{Appendix: Cluster images}
\vspace{1.0cm}
  \includegraphics[keepaspectratio=true,width=1.0\textwidth]{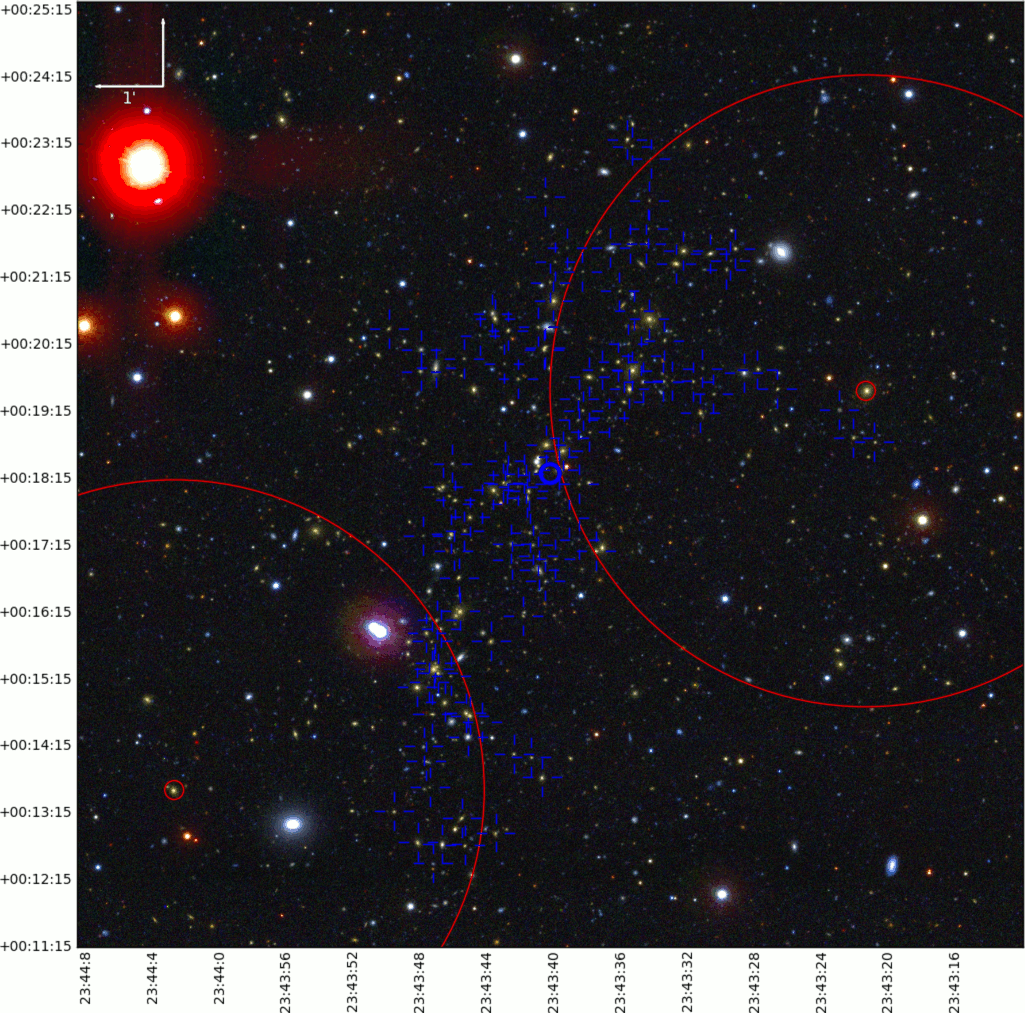}
\caption{Stripe 82 cluster {\it MGB J234341+00180.3} is an extended
  system detected between two {\tt maxBCG} clusters ({\it BCG
  J234322+00190.6} and {\it BCG J234403+00130.6}). For clarity, we
  have not plotted the Voronoi grid, but the cluster members are
  marked with blue cross-hairs. The {\tt maxBCG} clusters are shown in
  red, with the central positions noted by the two smaller circles,
  and the larger circles corresponding to radii of $1h^{-1}Mpc$ based
  on the photometrically-estimated cluster redshift from
  \citet{2007ApJ...660..239K}.
\label{filament}
}
\end{minipage}
\end{figure}

\newpage

\label{lastpage}
\end{document}